\documentclass[journal]{IEEEtran}
\IEEEoverridecommandlockouts

\usepackage{cite}
\usepackage{amssymb}
\usepackage{xcolor}
\usepackage{color} 
\usepackage{longtable}
\usepackage{booktabs}
\usepackage{multirow}
\usepackage{rotating}
\usepackage{float}
\usepackage{upgreek}
\usepackage{algorithmic}
\usepackage{algorithm}
\usepackage{tabularx}
\usepackage{amsmath,amsfonts}
\usepackage{array}
\usepackage[caption=false,font=footnotesize]{subfig}
\usepackage{textcomp}
\usepackage{stfloats}
\usepackage{url}
\usepackage{verbatim}
\usepackage{graphicx}
\usepackage[inline]{enumitem}
\usepackage{placeins}

\newcommand{\Propref}[1]{\textbf{Proposition~\ref{#1}}}

\usepackage{amsthm}
\newtheoremstyle{propstyle}
  {3pt}
  {3pt}
  {\normalfont}
  {}
  {\bfseries}
  {.}
  {0.5em}
  {}

\theoremstyle{propstyle}
\newtheorem{proposition}{Proposition}

\begin{document}
\setlength{\abovedisplayskip}{2pt}
\setlength{\belowdisplayskip}{2pt}
\title{Multi-Mode Pinching-Antenna-Based OFDM Systems: A Robust Design Perspective}

\author{Hao~Liang, Xidong~Mu,~\IEEEmembership{Member,~IEEE}, Mingzhe~Chen,~\IEEEmembership{Senior~Member,~IEEE}, Hyundong~Shin,~\IEEEmembership{Fellow,~IEEE},
and Michail~Matthaiou,~\IEEEmembership{Fellow,~IEEE}%
\thanks{H. Liang, X. Mu, and M. Matthaiou are with the Centre for Wireless
Innovation (CWI), Queen's University Belfast, Belfast BT3 9DT, U.K.
(e-mail: \{hliang06, x.mu, m.matthaiou\}@qub.ac.uk).}%
\thanks{M. Chen is with the Department of Electrical and Computer Engineering and the Frost Institute for Data Science and Computing, University of Miami, Coral Gables, FL 33146 USA (e-mail: mingzhe.chen@miami.edu).}%
\thanks{H. Shin is with the Department of Electronic Engineering, Kyung Hee
University, Yongin-si 17104, Republic of Korea (e-mail: hshin@khu.ac.kr).}}

\maketitle

\begin{abstract}
A robust transmission framework is proposed for a multi-mode
pinching-antenna (MPA)-based orthogonal frequency-division multiplexing
(OFDM) system under both channel state information (CSI) errors and PA
positioning errors. A physical model is first developed based on the
multi-mode coupled-mode equations (CMEs), which characterizes the
frequency-dependent guided-mode propagation and the coupling among guided
modes and PAs over the OFDM band. Based on this model, a worst-case
sum-rate maximization problem is formulated to jointly optimize the
baseband beamformer and the intended PA positions under CSI and PA positioning errors, subject to user
quality-of-service (QoS) and transmit-power constraints. To solve the
resulting non-convex problem, the worst-case CSI is exactly obtained for
any given PA positions by solving trust-region subproblems, while the PA
positioning errors are handled by grid-based and second-order
approximation-based scenario constructions. Then, an alternating
optimization (AO) algorithm is developed, where the baseband beamformer
and the intended PA positions are updated via the weighted minimum
mean-square error (WMMSE) method and successive convex approximation
(SCA), respectively. Both subproblems are solved efficiently by
exploiting the Karush--Kuhn--Tucker (KKT) conditions. Numerical results
show that the proposed robust design outperforms the non-robust and
conventional robust benchmarks with the same multi-mode architecture in
terms of both worst-case sum rate and QoS satisfaction. They also
demonstrate that the proposed robust design satisfies the QoS
requirements of all users more reliably than the baselines that consider
either CSI errors or PA positioning errors.
\end{abstract}


\section{Introduction}
Pinching-antenna (PA) technologies have recently emerged as promising solutions for achieving flexible and energy-efficient wireless communications [\citen{liu2026pinching},\citen{yang2025pinching}]. In particular, dielectric particles, namely PAs, are deployed along a long and low-loss waveguide, where their positions can be flexibly adjusted to bring the radiating points closer to users and establish unobstructed line-of-sight (LoS) links [\citen{xu2025rate},\citen{liu2025pinching}]. Such spatial reconfigurability can reduce free-space propagation loss, thereby enabling flexible reconfiguration of wireless channels \cite{ding2025flexible}. Recent studies have further investigated the physical principles of PA technologies, including electromagnetic coupling between PAs and dielectric waveguides \cite{wang2025modeling} and in-waveguide attenuation \cite{xu2026attenuation}.

Prior works on PA-based communication systems have mainly considered single-mode waveguides [\citen{xu2026attenuation},\citen{bereyhi2026mimo}], where a single waveguide provides only one effective spatial dimension and hence supports at most one independent data stream. Supporting simultaneous multi-user transmission may therefore require multiple waveguides, which increases the hardware cost and deployment complexity. To address this bottleneck, multi-mode PA (MPA) technologies were recently proposed, where multiple guided modes are excited within a single waveguide to transmit multiple independent data streams through mode-domain multiplexing \cite{xu2026multi}. In \cite{xu2026multi}, a physical model was developed to characterize mode-selective PA radiation, based on which mode-selection and mode-combining protocols were proposed in \cite{xu2026multi2}. In these works, mode-domain multiplexing relies on the relative phases among guided modes, which are characterized at a single frequency and under perfectly known channels and PA positions. Nevertheless, how these relative phases change across a wide bandwidth and under practical errors remains unexplored.

High-frequency communications represent a promising application of
PAs~\cite{fukuda2022pinching}, where wideband transmission is important
for exploiting the available spectrum. In~\cite{oikonomou2025ofdma}, the
frequency-selective channel caused by different PA propagation delays was
characterized, and an orthogonal frequency-division multiple access
(OFDMA) framework was developed to improve multi-user rate fairness. The
authors of~\cite{xiao2025frequency} developed a frequency-selective
channel model for OFDM-based PA systems, accounting for waveguide
dispersion, frequency-dependent attenuation, and PA coupling.
Moreover, the authors of~\cite{ouyang2026frequency} quantified the rate overestimation
caused by frequency-flat channel modeling and developed a
frequency-selective PA-position optimization method. However, these
studies consider a single guided mode. In MPA systems, different guided
modes experience different dispersion and different frequency-dependent
coupling to the PAs, so that the relative phases among modes vary across
subcarriers. Since the same PA positions are shared by all guided modes
and subcarriers, the existing single-mode models and position designs
cannot be directly applied.

Besides wideband modeling, the multiplexing gain of MPA systems relies on accurate knowledge of the effective channel, which is difficult to
obtain in practice. On one hand, CSI acquisition is challenging due
to limited pilot resources and channel estimation
noise~[\citen{zhu2025tutorial},\citen{xiao2025channel}]. On the other hand, finite
actuator resolution, mechanical drift, and vibration may cause the real
PA positions to deviate from their intended
values~\cite{xu2026generalized}. In MPA-OFDM systems, the effective
channel depends on the PA positions through both the free-space distances
and the in-waveguide propagation lengths. Since different guided modes
have different propagation constants, a small PA displacement shifts
their phases by different amounts across subcarriers, which cannot be
compensated by a common phase rotation. Hence, robust designs against
both CSI errors and PA positioning errors are essential. Existing robust
PA designs mainly considered CSI
errors~[\citen{sun2026robust},\citen{sun2026robustmultiuser}] or user-location
uncertainty~\cite{zeng2026robust,feng2026joint,zhao2026robust}, while
assuming that the PA positions are placed accurately. Although the impact
of PA positioning errors was analyzed
in~\cite{pakravan2026impact,jiang2026impact,yang2026movable}, these studies focus on
performance analysis rather than robust design.

Against the above background, the robust design of MPA-OFDM systems
under both CSI errors and PA positioning errors faces two main
challenges. First, the two types of errors cannot be handled separately.
Although they arise from independent sources, they act on the same
effective channel: a PA positioning error changes the waveguide-to-PA
response, which in turn determines how the CSI error affects the
received signal. Hence, the worst-case CSI error depends on the real PA
positions, and the two errors cannot be handled by simply combining a
CSI-robust design with a separate treatment of PA positioning errors.
Second, since the shared PA positions affect the channel nonlinearly, the
worst-case positioning error has no closed-form expression. To the best
of the authors' knowledge, there is no prior work addressing the robust
design of MPA-OFDM systems under both CSI errors and PA positioning
errors, which motivates this work.
 
Our main contributions are summarized as follows:
\begin{itemize}
    \item We develop a wideband MPA-OFDM channel model based on the
    multi-mode coupled-mode equations (CMEs), which captures the
    mode-dependent dispersion and the frequency-dependent coupling
    between guided modes and PAs. Based on this model, we characterize
    how CSI errors and PA positioning errors are coupled through the
    effective channel, and formulate a worst-case sum-rate maximization
    problem to jointly optimize the beamforming and intended PA positions.

    \item We derive the exact worst-case CSI error for any given PA-position realization via a low-dimensional complex trust-region subproblem. This position-dependent characterization explicitly captures the coupling between the two types of errors. The PA positioning errors are further represented by grid-based and second-order approximation-based scenario constructions.

    \item We develop a robust alternating optimization (AO) algorithm,
    where the beamforming and intended PA positions are updated via the
    weighted minimum mean-square error (WMMSE) method and successive
    convex approximation (SCA), respectively, with efficient
    Karush--Kuhn--Tucker (KKT)-based solutions and multi-start initialization.

    \item We evaluate the proposed designs under different CSI and PA positioning error levels. The results show that neglecting PA positioning errors significantly reduces the worst-case sum rate, while neglecting CSI errors mainly leads to QoS violations. By handling both errors jointly with multi-start initialization, the proposed designs achieve higher worst-case sum rates and more reliable QoS satisfaction than the non-robust and conventional robust multi-mode schemes.
\end{itemize}

The rest of this paper is organized as follows:
Section~\ref{sec:system_model} presents the system model and problem
formulation. Section~\ref{sec:robust_algorithm} develops the proposed
robust joint beamforming and PA-position optimization algorithms.
Section~\ref{sec:numerical} provides numerical results, and
Section~\ref{sec:conclusion} concludes the paper.

\textit{Notations:}
Scalars, vectors, and matrices are denoted by non-bold letters,
lowercase bold letters, and uppercase bold letters, respectively.
The operators $(\cdot)^T$, $(\cdot)^H$, $(\cdot)^*$, and
$(\cdot)^{\dagger}$ denote the transpose, Hermitian transpose, complex
conjugate, and Moore--Penrose pseudoinverse, respectively.
The symbols $\|\cdot\|_2$, $\|\cdot\|_F$, and $|\cdot|$ denote the
Euclidean norm, Frobenius norm, and absolute value (or set
cardinality), respectively.
The operators $\operatorname{diag}(\cdot)$, $\operatorname{Re}\{\cdot\}$,
$\operatorname{tr}(\cdot)$, $\exp(\cdot)$, and $\mathbb E[\cdot]$
denote the diagonal matrix construction, real part, trace, matrix
exponential, and expectation, respectively, and
$\mathrm j=\sqrt{-1}$ denotes the imaginary unit;
$\nabla_{\mathbf x}f$ and $\nabla_{\mathbf x}^2f$ denote the gradient
and Hessian of $f$ with respect to $\mathbf x$, and
$\mathbf X\succeq\mathbf 0$ ($\mathbf X\succ\mathbf 0$) indicates that
$\mathbf X$ is positive semidefinite (definite);
$\mathbf I_N$ denotes the $N\times N$ identity matrix,
$[\mathbf X]_{m,n}$ denotes the $(m,n)$-th entry of $\mathbf X$, and
$[\mathbf X]_{a:b,c:d}$ denotes the submatrix of $\mathbf X$ formed by
rows $a$ to $b$ and columns $c$ to $d$;
$\mathbb C$ and $\mathbb R$ denote the sets of complex and real
numbers, respectively, $\prod$ denotes the Cartesian product of sets,
and $\mathcal{CN}(0,\sigma^2)$ denotes the circularly symmetric complex
Gaussian distribution with zero mean and variance $\sigma^2$.

\section{System Model and Problem Formulation}
\label{sec:system_model}
This section presents the MPA-based OFDM system model and formulates
the corresponding robust design problem. 

\subsection{Wideband MPA Model}
\label{subsec:wideband_model}
\begin{figure}[t]
	\centering	\includegraphics[width=\linewidth]{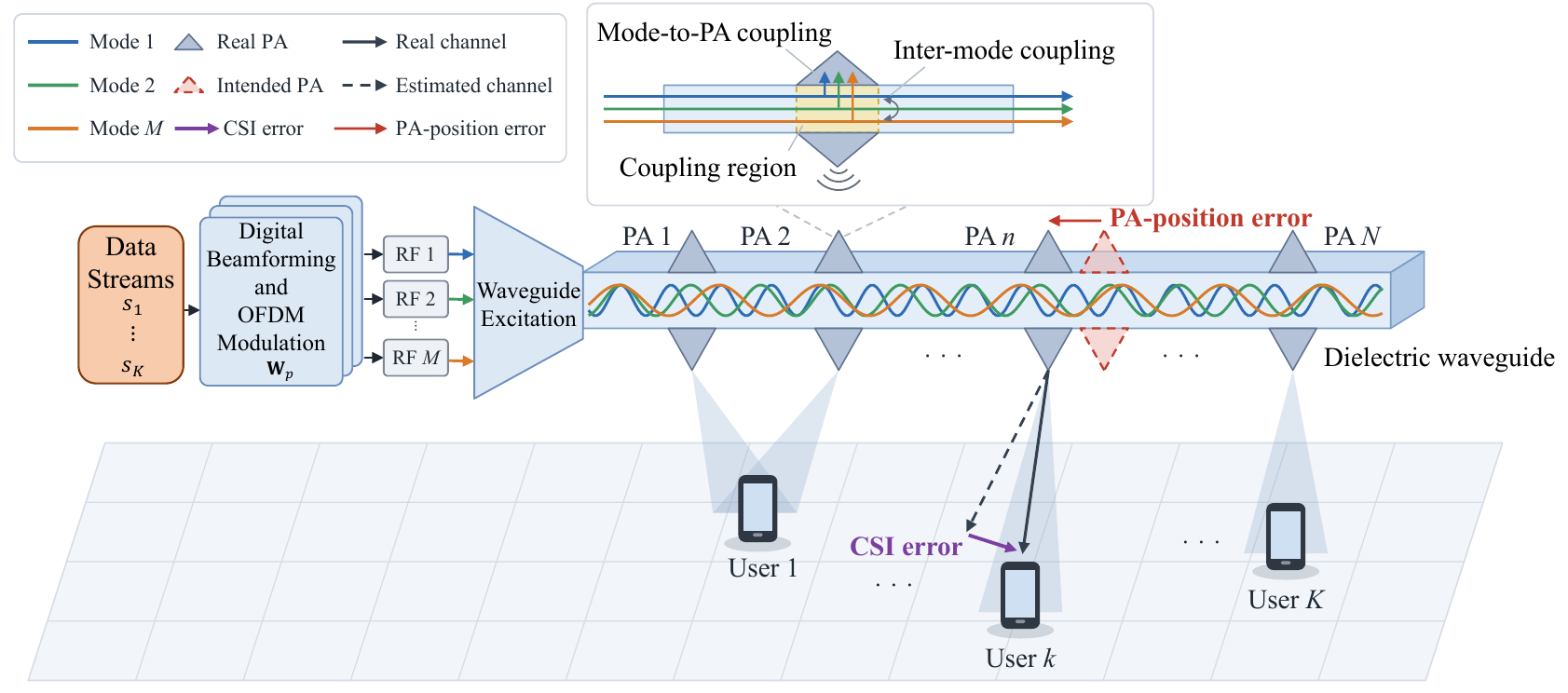}
	\caption{\centering The proposed MPA-based OFDM system.}
    \vspace{-1mm}
	\label{fig_1}
    \vspace{-1mm}
\end{figure}

We first provide a physical model for an MPA-based OFDM system with multiple PAs, which captures the frequency-dependent signal propagation over a set of $P$ subcarriers within an $M$-mode waveguide. As illustrated in Fig.~\ref{fig_1}, the OFDM subcarriers and guided modes are indexed by $\mathcal P=\{1,\ldots,P\}$ and $\mathcal M=\{1,\ldots,M\}$, respectively. The signal over each subcarrier $p\in\mathcal P$ is carried by the $M$ guided modes, which are excited by $M$ dedicated feeds at the waveguide. Consider a set of $N$ PAs, indexed by
$\mathcal N=\{1,\ldots,N\}$, which are sequentially located along the waveguide at the ordered positions
$0<x_1<x_2<\cdots<x_N<L_{\mathrm{wg}}$, where $L_{\mathrm{wg}}$ denotes the length of the waveguide. For guided mode $m$ over subcarrier $p$, let $\gamma_{m,p}\triangleq\alpha_{m,p}+\mathrm j\beta_{m,p}$ denote the complex
propagation constant. Here, $\alpha_{m,p}$ is the waveguide attenuation
coefficient in nepers per meter, while $\beta_{m,p}$ is the phase constant given by
\begin{equation}
	\beta_{m,p}=\frac{2\pi f_p}{c}n_{\mathrm{eff},m}(f_p),
	\label{eq:mode_phase_constant}
\end{equation}
where $n_{{\rm eff},m}(f_p)$ denotes the frequency-dependent
effective phase index of guided mode $m$, while $f_p$ and $c$
denote the subcarrier frequency and the speed of light,
respectively.

For PA $n$ operating over subcarrier $p$, let
$A_{n,m,p}(\zeta)$ and $B_{n,p}(\zeta)$ denote the complex
amplitudes of guided mode $m$ and the PA eigenmode,
respectively, within the local coupling region
$\zeta\in[0,L_n]$, where $L_n$ is the effective coupling length.
Each PA supports a single eigenmode over the considered OFDM
band, which electromagnetically couples to the $M$ guided modes
for signal radiation. To characterize the complete local
multi-mode interaction, define
$\boldsymbol{\psi}_{n,p}(\zeta)
=[A_{n,1,p}(\zeta),\ldots,A_{n,M,p}(\zeta),B_{n,p}(\zeta)]^T\in\mathbb C^{(M+1)\times1}$.
The corresponding multi-mode CME
can be written as
\begin{equation}
    \frac{{\rm d}\boldsymbol{\psi}_{n,p}(\zeta)}
    {{\rm d}\zeta}
    =
    -\mathrm j \mathbf C_{n,p}
    \boldsymbol{\psi}_{n,p}(\zeta),
    \label{eq:full_cmt_state}
\end{equation}
where
\begin{equation}
    \mathbf C_{n,p}
    =
    \begin{bmatrix}
        \mathbf C^{\rm w}_{n,p}
        &
        \boldsymbol{\kappa}_{n,p}
        \\
        \boldsymbol{\kappa}_{n,p}^{H}
        &
        \beta_{{\rm PA},n,p}
    \end{bmatrix}\in\mathbb C^{(M+1)\times(M+1)}.
    \label{eq:full_cmt_matrix}
\end{equation}
Here, $\beta_{{\rm PA},n,p}$ denotes the propagation constant
of the PA eigenmode and
$\boldsymbol{\kappa}_{n,p}
=[\kappa_{n,1,p},\ldots,\kappa_{n,M,p}]^T \in\mathbb C^{M\times1}$
collects the coupling coefficients between the guided modes
and the PA eigenmode. The guided-mode block
$\mathbf C^{\rm w}_{n,p}\in\mathbb C^{M\times M}$ is given by
\begin{equation}
    [\mathbf C^{\rm w}_{n,p}]_{m,m'}
    =
    \begin{cases}
        \beta_{m,p}, & m=m',\\
        \kappa^{\rm ww}_{n,m,m',p}, & m\neq m',
    \end{cases}
    \label{eq:Hww}
\end{equation}
where $\kappa^{\rm ww}_{n,m,m',p}$ denotes the coupling
coefficient between guided modes $m$ and $m'$ within the PA
coupling region. For reciprocal and locally lossless coupling,
$\mathbf C_{n,p}$ is Hermitian.

Assuming that the coupling parameters are approximately
constant over the short coupling region, we denote the local propagation
matrix as
$\mathbf U_{n,p}=\exp\left(-\mathrm j\mathbf C_{n,p}L_n\right)
\in\mathbb C^{(M+1)\times(M+1)}$.
The guided-mode transfer matrix and the guided-mode-to-PA
response are obtained from $\mathbf U_{n,p}$ as
$\mathbf D_{n,p}=[\mathbf U_{n,p}]_{1:M,1:M}\in\mathbb C^{M\times M}$
and
$\mathbf r_{n,p}^{H}=[\mathbf U_{n,p}]_{M+1,1:M}\in\mathbb C^{1\times M}$,
respectively. Since the PA eigenmode is passively excited by
the guided modes, we set $B_{n,p}(0)=0$. Hence,
$\mathbf D_{n,p}$ characterizes the guided-mode signal
transmitted through the local coupling region, while
$\mathbf r_{n,p}^{H}$ characterizes the signal coupled to
PA $n$ for radiation. We then define the de-embedded
responses
$\overline{\mathbf D}_{n,p}\in\mathbb C^{M\times M}$ and
$\overline{\mathbf r}_{n,p}^{H}\in\mathbb C^{1\times M}$ as\footnote{With this de-embedding, if all local coupling coefficients
vanish, we have
$\overline{\mathbf D}_{n,p}=\mathbf I_M$ and
$\overline{\mathbf r}_{n,p}^{H}=\mathbf 0^{H}$, such that the
local PA region introduces no additional effect beyond the
waveguide propagation already captured by the propagation
matrices.}
\begin{align}
    \overline{\mathbf D}_{n,p}
    &=
    {\rm diag}
    \left(
    e^{\mathrm j\beta_{1,p}L_n},\ldots,
    e^{\mathrm j\beta_{M,p}L_n}
    \right)
    \mathbf D_{n,p},
    \label{eq:C_bar}\\
    \overline{\mathbf r}_{n,p}^{H}
    &=
    e^{\mathrm j\beta_{{\rm PA},n,p}L_n}
    \mathbf r_{n,p}^{H}.
    \label{eq:r_bar}
\end{align}

Let $x_0\triangleq0$ denote the position of the waveguide input. For
$n=0,\ldots,N-1$, the propagation from $x_n$ to PA $n+1$ over
subcarrier $p$ is characterized by the $M\times M$ diagonal matrix
\begin{equation}
    \mathbf P_{n,p}
    =
    {\rm diag}
    \left(
    e^{-\gamma_{1,p}(x_{n+1}-x_n)},\ldots,
    e^{-\gamma_{M,p}(x_{n+1}-x_n)}
    \right).
    \label{eq:Pn}
\end{equation}

The response from the $M$ waveguide inputs to PA $1$ is then
$\mathbf g_{1,p}^{H}
=\overline{\mathbf r}_{1,p}^{H}\mathbf P_{0,p}\in\mathbb C^{1\times M}$.
For PA $n\ge2$, the signal experiences the local coupling and
waveguide propagation associated with all preceding PAs.
Accordingly, for $n=2,\ldots,N$,
\begin{align}
    \mathbf g_{n,p}^{H}
    =
    \overline{\mathbf r}_{n,p}^{H}
    \mathbf P_{n-1,p}
    \overline{\mathbf D}_{n-1,p}
    \cdots
    \mathbf P_{1,p}
    \overline{\mathbf D}_{1,p}
    \mathbf P_{0,p}.
    \label{eq:g_full}
\end{align}

By stacking the responses of all PAs, the waveguide-to-PA
channel matrix over subcarrier $p$ is given by
\begin{equation}
   \mathbf G_p=[\mathbf g_{1,p},\ldots,\mathbf g_{N,p}]^H
    \in\mathbb C^{N\times M}.
    \label{eq:G_full}
\end{equation}

\subsection{Signal Model}
\label{subsec:signal_model}

We consider $K$ single-antenna communication users, indexed by
$\mathcal K=\{1,\ldots,K\}$. User $k$ is located at $(a_k,b_k,0)$ and the waveguide is deployed
at height $h_{\rm PA}$. The data-symbol vector over subcarrier
$p$ is denoted by
$\mathbf s_p=[s_{1,p},\ldots,s_{K,p}]^T\in\mathbb C^{K\times1}$,
which satisfies
$\mathbb E[\mathbf s_p\mathbf s_p^H]=\mathbf I_K$.
A digital baseband beamformer
$\mathbf W_p=[\mathbf w_{1,p},\ldots,\mathbf w_{K,p}]
\in\mathbb C^{M\times K}$
maps the data symbols to the $M$ modal excitation signals.
Accordingly, the signals fed into the $M$ guided modes over
subcarrier $p$ are given by
$\mathbf W_p\mathbf s_p$, and the signals radiated
by the $N$ PAs can be expressed as $\mathbf G_p\mathbf W_p\mathbf s_p \in\mathbb C^{N \times 1}$.

In practice, the real PA positions may deviate from the
positions specified by the system controller due to the finite
positioning accuracy of the PA actuation mechanism
\cite{xu2026generalized}. Let
$\widehat{\mathbf x}
=[\widehat x_1,\ldots,\widehat x_N]^T \in\mathbb R^{N\times1}$
denote the intended PA-position vector. The real PA-position
vector is modeled as
\begin{equation}
    \mathbf x
    =
    \widehat{\mathbf x}
    +
    \Delta\mathbf x,
    \label{eq:position_error}
\end{equation}
where
$\Delta\mathbf x=[\Delta x_1,\ldots,\Delta x_N]^T \in\mathbb R^{N \times 1}$
denotes the positioning error. We assume
$\Delta\mathbf x\in\mathcal U_x$, where
\begin{equation}
    \mathcal U_x
    =
    \left\{
    \Delta\mathbf x\in\mathbb R^N:
    |\Delta x_n|\le\delta_n,\;
    n\in\mathcal N
    \right\}.
    \label{eq:position_error_set}
\end{equation}
Here, $\delta_n$ denotes the maximum positioning error of PA
$n$. For homogeneous positioning accuracy, we set
$\delta_n=\delta_x$, $\forall n\in\mathcal N$.
Accordingly, under a positioning-error realization
$\Delta\mathbf x$, the waveguide-to-PA response derived in
Section~\ref{subsec:wideband_model} is evaluated at the real
PA positions
$\mathbf x=\widehat{\mathbf x}+\Delta\mathbf x$ and is denoted
by $\mathbf G_p(\mathbf x)$.

For a given PA-position vector $\mathbf x$, the estimated channel
between PA $n$ and user $k$ over subcarrier $p$ is modeled by the
spherical-wave LoS response
\begin{equation}
    \hat h_{n,k,p}(\mathbf x)
    =
    \frac{c}{4\pi f_p r_{n,k}(\mathbf x)}
    \exp\left(
    \mathrm j\frac{2\pi f_p}{c}r_{n,k}(\mathbf x)
    \right),
    \label{eq:los_channel}
\end{equation}
where $r_{n,k}(\mathbf x)=\sqrt{(x_n-a_k)^2+b_k^2+h_{\rm PA}^2}$
denotes the distance between PA $n$ and user $k$. Define $\widehat{\mathbf h}_{k,p}(\mathbf x)
\triangleq[\hat h_{1,k,p}(\mathbf x),\ldots,
\hat h_{N,k,p}(\mathbf x)]^T\in\mathbb C^{N\times1}$. Based on the available channel model, the transmitter can evaluate
$\widehat{\mathbf h}_{k,p}(\cdot)$ for any given PA positions.
However, since $\Delta\mathbf x$ is unknown, the value of
$\widehat{\mathbf h}_{k,p}(\mathbf x)$ at the real PA positions is
not known at the transmitter. Moreover, due to imperfect channel estimation, the real channel vector is modeled as
\begin{equation}
    \mathbf h_{k,p}(\widehat{\mathbf x}+\Delta\mathbf x)
    =
    \widehat{\mathbf h}_{k,p}
    \left(
    \widehat{\mathbf x}+\Delta\mathbf x
    \right)
    +
    \Delta\mathbf h_{k,p},
    \label{eq:real_channel}
\end{equation}
where $\Delta\mathbf h_{k,p}$ denotes the CSI error. We adopt
a deterministic norm-bounded error model with
$\Delta\mathbf h_{k,p}\in\mathcal U_{k,p}^{\rm h}$, where
\begin{equation}
    \mathcal U_{k,p}^{\rm h}
    =
    \left\{
    \Delta\mathbf h_{k,p}\in\mathbb C^N:
    \left\|
    \Delta\mathbf h_{k,p}
    \right\|_2
    \le
    \epsilon_{k,p}
    \right\}.
    \label{eq:csi_error_set}
\end{equation}
Here, $\epsilon_{k,p}$ denotes the CSI error radius of
user $k$ over subcarrier $p$.

Based on the real PA positions and channels, the
received signal at user $k$ over subcarrier $p$ is given by
\begin{align}
    y_{k,p}
    &={}
    \mathbf h_{k,p}(\mathbf x)^{H}
    \mathbf G_p(\mathbf x)
    \mathbf w_{k,p}s_{k,p}\notag \\ 
    &+
    \sum_{\substack{j\in\mathcal K, j\neq k}}
    \mathbf h_{k,p}(\mathbf x)^{H}
    \mathbf G_p(\mathbf x)
    \mathbf w_{j,p}s_{j,p}
    +
    n_{k,p},
    \label{eq:received_signal}
\end{align}
where
$n_{k,p}\sim\mathcal{CN}(0,\sigma_{k,p}^2)$ denotes the
additive white Gaussian noise. The resulting signal-to-interference-plus-noise ratio (SINR) of user $k$
over subcarrier $p$ can be expressed as
\begin{equation}
    \Gamma_{k,p}
    =
    \frac{
    \left|
    \mathbf h_{k,p}(\mathbf x)^{H}
    \mathbf G_p(\mathbf x)
    \mathbf w_{k,p}
    \right|^2
    }{
    \displaystyle
    \sum_{\substack{j\in\mathcal K, j\neq k}}
    \left|
    \mathbf h_{k,p}(\mathbf x)^{H}
    \mathbf G_p(\mathbf x)
    \mathbf w_{j,p}
    \right|^2
    +
    \sigma_{k,p}^2
    }.
    \label{eq:sinr}
\end{equation}

Accordingly, the achievable rate of user $k$ across all
subcarriers is given by
\begin{equation}
    R_k
    =
    \frac{B}{P+L_{\rm CP}}
    \sum_{p\in\mathcal P}
    \log_2
    \left(
    1+\Gamma_{k,p}
    \right),
    \label{eq:user_rate}
\end{equation}
where $B$ and $L_{\rm CP}$ denote the system bandwidth and the
CP length, respectively.


\subsection{Problem Formulation}
\label{subsec:problem_formulation}

Define $\mathbf W\triangleq\{\mathbf W_p\}_{p\in\mathcal P}$ and
$\Delta\mathbf h_k\triangleq
\{\Delta\mathbf h_{k,p}\}_{p\in\mathcal P}$, with
$\mathcal U_k^{\rm h}\triangleq
\prod_{p\in\mathcal P}\mathcal U_{k,p}^{\rm h}$.
We jointly optimize the baseband beamforming and the
intended PA positions to maximize the worst-case system sum
rate under both PA-position and CSI errors. The
resulting problem is formulated as
\begin{subequations}
\label{prob:robust_design}
\begin{align}
    \max_{\mathbf W,\widehat{\mathbf x}}
    &
    \min_{\Delta\mathbf x\in\mathcal U_x}
    \sum_{k\in\mathcal K}
    \min_{\Delta\mathbf h_k\in\mathcal U_k^{\rm h}}
    R_k
    \left(
    \mathbf W,
    \widehat{\mathbf x}+\Delta\mathbf x,
    \Delta\mathbf h_k
    \right)
    \label{prob:robust_obj}\\
    \mathrm{s.t.}&
    \min_{\Delta\mathbf x\in\mathcal U_x}
    \min_{\Delta\mathbf h_k\in\mathcal U_k^{\rm h}}
    R_k
    \left(
    \mathbf W,
    \widehat{\mathbf x}+\Delta\mathbf x,
    \Delta\mathbf h_k
    \right)
    \ge R_k^{\min},
     \forall k\in\mathcal K,
    \label{prob:robust_qos}\\
    &
    \|\mathbf W_p\|_F^2
    \le P_{p,\max},
    \quad p\in\mathcal P,
    \label{prob:subcarrier_power}\\
    &
    x_{\min}+\delta_x
    \le \widehat x_n
    \le x_{\max}-\delta_x,
    \quad n\in\mathcal N,
    \label{prob:position_boundary}\\
    &
    \widehat x_{n+1}-\widehat x_n
    \ge
    \Delta_{\min}+2\delta_x,
    n=1,\ldots,N-1,
    \label{prob:position_spacing}
\end{align}
\end{subequations}
where $P_{p,\max}$ denotes the
per-subcarrier transmit-power limits. In this work, we adopt the equal per-subcarrier power limit
$P_{p,\max}=P_{\max}/P$, $\forall p\in\mathcal P$, where $P_{\max}$ denotes the total transmit power.
Therefore, the total transmit-power constraint is automatically
satisfied by the per-subcarrier constraints and is omitted
hereafter, while
$R_k^{\min}$ denotes the minimum rate requirement of user $k$.
Moreover, $[x_{\min},x_{\max}]$ specifies the feasible PA
deployment region and $\Delta_{\min}$ denotes the minimum
spacing between two adjacent PAs. Constraints
\eqref{prob:position_boundary} and
\eqref{prob:position_spacing} ensure that the real PA
positions remain physically feasible for every
$\Delta\mathbf x\in\mathcal U_x$.

Problem~\eqref{prob:robust_design} is highly non-convex due to the
coupled desired-signal and inter-user-interference terms and the nested
worst-case optimization over the PA-position and CSI errors. In
particular, the two types of errors cannot be handled independently. 
Although $\Delta\mathbf h_{k,p}$ and $\Delta\mathbf x$ arise from
independent sources, the CSI error affects the received signal only
through $\mathbf G_p(\mathbf x)$, which depends on the real PA
positions. Since different guided modes have different propagation
constants, a PA positioning error further changes the phases of different
guided modes in $\mathbf G_p(\mathbf x)$ by different amounts. Hence,
the worst-case CSI error varies with the PA positioning error, and the two
types of errors have to be jointly considered. Moreover, the same
PA-position vector is shared by all guided modes, subcarriers, and
users. Therefore, directly obtaining the globally optimal solution of
Problem~\eqref{prob:robust_design} is generally intractable.

\section{Robust Joint Beamforming and PA-Position Optimization}
\label{sec:robust_algorithm}

To solve Problem~\eqref{prob:robust_design}, we develop an
AO framework to jointly optimize the
baseband beamforming and the intended PA positions.
To capture the coupling between the two types of errors,
the CSI error is handled conditioned on the PA positions.
Specifically, the worst-case rate under the norm-bounded CSI
error is first characterized for any given PA-position vector.
The remaining PA positioning error is then approximated
using a finite set of representative position-error scenarios.
Based on this formulation, the beamforming and intended PA positions are alternately updated via WMMSE and SCA, respectively.

\subsection{Worst-Case CSI Characterization}
\label{subsec:exact_csi}

For an arbitrary real PA-position vector $\mathbf x$, the CSI
error in \eqref{eq:csi_error_set} can be equivalently written as
$\Delta\mathbf h_{k,p}=\epsilon_{k,p}\mathbf q_{k,p}$ with
$\|\mathbf q_{k,p}\|_2\le1$, where $\mathbf q_{k,p}\in\mathbb C^N$
denotes the normalized CSI-error vector.

Let $\mathbf \Omega_p \triangleq (\mathbf W_p, \mathbf x)$. For user $k$, data stream $j$, and subcarrier $p$, define
\begin{align}
    \mu_{k,j,p}(\mathbf \Omega_p)
    &\triangleq
    \frac{
    \widehat{\mathbf h}_{k,p}^{H}(\mathbf x)
    \mathbf G_p(\mathbf x)
    \mathbf w_{j,p}}
    {\sigma_{k,p}},
    \label{eq:mu_def}\\
    \boldsymbol{\nu}_{k,j,p}(\mathbf \Omega_p)
    &\triangleq
    \frac{
    \epsilon_{k,p}
    \mathbf G_p(\mathbf x)
    \mathbf w_{j,p}}
    {\sigma_{k,p}}.
    \label{eq:nu_def}
\end{align}
Then, the normalized effective channel associated with
$\mathbf w_{j,p}$ can be expressed as
\begin{equation}
    \frac{
    \mathbf h_{k,p}^{H}
    \mathbf G_p(\mathbf x)
    \mathbf w_{j,p}}
    {\sigma_{k,p}}
    =
    \mu_{k,j,p}(\mathbf \Omega_p)
    +
    \mathbf q_{k,p}^{H}
    \boldsymbol{\nu}_{k,j,p}(\mathbf \Omega_p).
    \label{eq:uncertain_scalar_channel}
\end{equation}

For fixed $\mathbf \Omega_p$, minimizing the achievable rate of a user-subcarrier pair is equivalent to minimizing its SINR. Hence, the worst-case SINR of user $k$ over subcarrier $p$ is given by
{
\begin{equation}
\begin{aligned}
    \min_{\|\mathbf q_{k,p}\|_2\le1}
    \frac{
    \left|
    \mu_{k,k,p}(\mathbf \Omega_p)
    +
    \mathbf q_{k,p}^{H}
    \boldsymbol{\nu}_{k,k,p}(\mathbf \Omega_p)
    \right|^2
    }{
    1+
    \displaystyle
    \sum_{\substack{j\in\mathcal K\\j\neq k}}
    \left|
    \mu_{k,j,p}(\mathbf \Omega_p)
    +
    \mathbf q_{k,p}^{H}
    \boldsymbol{\nu}_{k,j,p}(\mathbf \Omega_p)
    \right|^2
    }.
\end{aligned}
\label{eq:wc_sinr}
\end{equation}
}
Let $ \Gamma_{k,p}^{\rm wc}(\mathbf \Omega_p)$ denote the optimal value of Problem~\eqref{eq:wc_sinr}, which is characterized exactly by the following proposition.

\begin{proposition}[\textbf{Worst-Case CSI}]
\label{Proposition_CSI}
For fixed $\mathbf\Omega_p$ and any $\tau\ge0$, define
\par{\small
\begin{align}
\label{eq:dinkelbach_function}
    \phi_{k,p}(\tau;\mathbf\Omega_p)
    &\triangleq
    \min_{\|\mathbf q_{k,p}\|_2\le1}
    \Big\{  
    \left|
    \mu_{k,k,p}(\mathbf\Omega_p)
    +
    \mathbf q_{k,p}^{H}
    \boldsymbol{\nu}_{k,k,p}(\mathbf\Omega_p)
    \right|^2 \notag
    \\
   & -\tau[1+\!\!\!\!\!\sum_{\substack{j\in\mathcal K, j\neq k}}
    \left|
    \mu_{k,j,p}(\mathbf\Omega_p)
    +\mathbf q_{k,p}^{H}\boldsymbol{\nu}_{k,j,p}(\mathbf\Omega_p)
    \right|^2]\Big\}.
\end{align}}
\hspace*{-\parindent}~\,Then, $\phi_{k,p}(\tau;\mathbf\Omega_p)$ is strictly decreasing
in $\tau$, and $\Gamma_{k,p}^{\rm wc}(\mathbf\Omega_p)$ is
the unique nonnegative root of
$\phi_{k,p}(\tau;\mathbf\Omega_p)=0$. Moreover, at least one optimal
CSI-error vector can be chosen from
${\rm span}\{\boldsymbol{\nu}_{k,1,p},\ldots,\boldsymbol{\nu}_{k,K,p}\}$,
whose dimension is at most $\min\{N,M,K\}$. Consequently, for any fixed $\tau$, the minimization over
$\mathbf q_{k,p}$ in \eqref{eq:dinkelbach_function} can be
solved globally as a low-dimensional complex trust-region
subproblem.
\end{proposition}

\emph{Proof: See Appendix~\ref{app:proof_adversary}.}

\Propref{Proposition_CSI} eliminates the CSI error for any
fixed $\mathbf\Omega_p$. Since the CSI error sets are independent across
subcarriers, the corresponding worst-case rate of
user $k$ for fixed $\mathbf W$ and $\mathbf x$ is given by
\begin{equation}
    R_k^{\rm wc}(\mathbf W,\mathbf x)
    =
    \frac{B}{P+L_{\rm CP}}
    \sum_{p\in\mathcal P}
    \log_2
    \left(
    1+
    \Gamma_{k,p}^{\rm wc}(\mathbf W_p,\mathbf x)
    \right).
    \label{eq:wc_user_rate}
\end{equation}

Using the worst-case CSI characterization in
\Propref{Proposition_CSI}, Problem~\eqref{prob:robust_design} can be
rewritten as
\begin{subequations}
\label{prob:continuous_position_robust}
\begin{align}
    \max_{\mathbf W,\widehat{\mathbf x}}
    &\min_{\Delta\mathbf x\in\mathcal U_x}
    \sum_{k\in\mathcal K}
    R_k^{\rm wc}
    \left(
    \mathbf W,
    \widehat{\mathbf x}+\Delta\mathbf x
    \right)
    \label{prob:continuous_position_obj}\\
    \mathrm{s.t.}
    &\min_{\Delta\mathbf x\in\mathcal U_x}
    R_k^{\rm wc}
    \left(
    \mathbf W,
    \widehat{\mathbf x}+\Delta\mathbf x
    \right)
    \ge R_k^{\min},
    \forall k\in\mathcal K,
    \label{prob:continuous_position_qos}\\  
    &\eqref{prob:subcarrier_power}, ~ \eqref{prob:position_boundary}, ~
    \eqref{prob:position_spacing}.
\end{align}
\end{subequations}

The remaining difficulty in
Problem~\eqref{prob:continuous_position_robust}
lies in the continuous minimizations over
$\Delta\mathbf x\in\mathcal U_x$.
Since both
$\widehat{\mathbf h}_{k,p}(\mathbf x)$ and
$\mathbf G_p(\mathbf x)$ depend nonlinearly on the real PA
positions, these minimizations are difficult to handle directly.
We therefore represent the position error by a finite
collection of representative position-error realizations.

\subsection{Position-Error Scenario Construction}
\label{subsec:position_scenario_construction}
Let $\mathcal S$ denote the index set of the finite
position-error scenarios used to characterize the worst-case
PA positioning error.
For each $s\in\mathcal S$, let
$\Delta\mathbf x^{(s)}\in\mathcal U_x$
denote the associated position-error realization.
Given an intended PA-position vector
$\widehat{\mathbf x}$, the resulting real PA-position vector
under scenario $s$ is given by
$\mathbf x_s=\widehat{\mathbf x}+\Delta\mathbf x^{(s)}$, $s\in\mathcal S$.

The continuous robust objective and QoS constraints in
Problem~\eqref{prob:continuous_position_robust} are then approximated
using the selected position-error scenarios. Specifically, the robust
objective in \eqref{prob:continuous_position_obj} is approximated by the
finite-scenario robust sum rate
\begin{equation}
    J_{\mathcal S}
    \left(
        \mathbf W,\widehat{\mathbf x}
    \right)
    \triangleq
    \min_{s\in\mathcal S}
    \sum_{k\in\mathcal K}
    R_k^{\rm wc}
    \left(
        \mathbf W,
        \widehat{\mathbf x}
        +
        \Delta\mathbf x^{(s)}
    \right),
    \label{eq:finite_scenario_objective}
\end{equation}
and the QoS constraints in \eqref{prob:continuous_position_qos} are
approximated by
\begin{equation}
    R_k^{\rm wc}
    \left(
        \mathbf W,
        \mathbf x_s
    \right)
    \ge
    R_k^{\min},
    \qquad
    k\in\mathcal K,\;
    s\in\mathcal S.
    \label{eq:finite_position_qos}
\end{equation}
Since $\Delta\mathbf x^{(s)}\in\mathcal U_x$, $\forall s\in\mathcal S$,
$J_{\mathcal S}(\mathbf W,\widehat{\mathbf x})$ is an upper bound of the
continuous worst-case objective in \eqref{prob:continuous_position_obj},
and its accuracy depends on how well $\mathcal S$ captures the critical
position errors. In the following, we consider two different constructions of
$\mathcal S$. The first one is based on Cartesian discretization,
whereas the second one constructs the scenario set based on a second-order approximation over the continuous position-error region.
\subsubsection{Grid-Based Scenario Construction}
\label{subsubsec:grid_scenario}
The first strategy uniformly discretizes each component of
$\mathcal U_x$. Let $Q\ge2$ denote the number of sampled
position-error values along each dimension, and define
\begin{equation}
    \mathcal D_n(Q)
    \triangleq
    \left\{
    -\delta_n
    +
    \frac{2(\ell-1)\delta_n}{Q-1}
    :
    \ell=1,\ldots,Q
    \right\}, n\in\mathcal N .
    \label{eq:position_grid_1d}
\end{equation}
The resulting Cartesian grid is
\begin{equation}
    \mathcal U_{x}^{(Q)}
    \triangleq
    \mathcal D_1(Q)
    \times\cdots\times
    \mathcal D_N(Q)
    =
    \left\{
    \Delta\mathbf x_{\rm G}^{(s)}
    :
    s\in\mathcal S_{\rm G}
    \right\}
    \subseteq
    \mathcal U_x ,
    \label{eq:position_grid}
\end{equation}
where
$\mathcal S_{\rm G}
\triangleq
\{1,\ldots,Q^N\}$
and $|\mathcal S_{\rm G}|=Q^N$. For this construction, we set
$\mathcal S=\mathcal S_{\rm G}$ and
$\Delta\mathbf x^{(s)}=\Delta\mathbf x_{\rm G}^{(s)}$.

\subsubsection{Second-Order Approximation-Based Construction}
\label{subsubsec:trueb_scenario}

Instead of discretizing $\mathcal U_x$, the second strategy
constructs the position-error scenarios directly based on a
second-order approximation of the worst-case rates.
For fixed $\mathbf W$ and $\widehat{\mathbf x}$, define
$\mathcal R_k(\Delta\mathbf x)\triangleq
R_k^{\rm wc}(\mathbf W,\widehat{\mathbf x}+\Delta\mathbf x)$,
$k\in\mathcal K$, and
$\mathcal R_0(\Delta\mathbf x)\triangleq
\sum_{k\in\mathcal K}\mathcal R_k(\Delta\mathbf x)$.
Here, $\mathcal R_0$ corresponds to the robust sum-rate objective in
\eqref{prob:continuous_position_obj}, whereas $\mathcal R_k$ is associated
with the QoS constraint of user $k$ in
\eqref{prob:continuous_position_qos}.

For $l \in\{0\}\cup\mathcal K$, a second-order approximation
of $\mathcal R_l(\Delta\mathbf x)$ around
$\Delta\mathbf x=\mathbf 0$ is given by
\begin{equation}
\begin{aligned}
    \mathcal R_l(\Delta\mathbf x)
    \approx\;&
    \mathcal R_l(\mathbf 0)
    +
    \mathbf a_l^{T}\Delta\mathbf x
    +
    \frac{1}{2}
    \Delta\mathbf x^{T}
    \mathbf F_l
    \Delta\mathbf x ,
\end{aligned}
\label{eq:second_order_position_model}
\end{equation}
where
$\mathbf a_l\!\triangleq \!
\left.
\nabla_{\Delta\mathbf x}\mathcal R_l(\Delta\mathbf x)
\right|_{\Delta\mathbf x=\mathbf 0}$
and
$\mathbf F_l\triangleq
\left.
\nabla_{\Delta\mathbf x}^{2}\mathcal R_l(\Delta\mathbf x)
\right|_{\Delta\mathbf x=\mathbf 0}$. 
Since $\mathcal R_l$ is obtained from the worst-case CSI problem in \Propref{Proposition_CSI}, we assume that
the worst-case CSI error is locally unique and
its KKT system is nonsingular around
$\Delta\mathbf x=\mathbf 0$, ensuring the local
twice differentiability of $\mathcal R_l$.
When $\epsilon_{k,p}=0$, the corresponding inner
CSI minimization is unnecessary, and the rate
derivatives are evaluated directly.

Based on \eqref{eq:second_order_position_model}, for each
$l\in\{0\}\cup\mathcal K$, the candidate worst-case position errors are
obtained from the box-constrained quadratic problem
\begin{equation}
    \underset{\Delta\mathbf x\in\mathcal U_x}{\operatorname{minimize}}
    \quad
    \mathcal R_l(\mathbf 0)
    +
    \mathbf a_l^{T}\Delta\mathbf x
    +
    \frac{1}{2}
    \Delta\mathbf x^{T}
    \mathbf F_l
    \Delta\mathbf x ,
\label{prob:second_order_system}
\end{equation}
where $l=0$ corresponds to the system objective and $l=k\in\mathcal K$
corresponds to the QoS constraint of user $k$. The $K+1$ quadratic
problems in \eqref{prob:second_order_system} can be solved by enumerating the $3^N$
active-bound patterns of the position-error coordinates, which only
involves algebraic operations on
\eqref{eq:second_order_position_model}. We retain the local minimizers
and remove duplicates.

Let $\left\{ \Delta\mathbf x_{\rm B}^{(s)}:s\in\mathcal S_{\rm B}\right\}\subseteq\mathcal U_x$ denote the union of the retained solutions obtained from
\eqref{prob:second_order_system}. For the second-order construction, we
therefore set $\mathcal S=\mathcal S_{\rm B}$ and
$\Delta\mathbf x^{(s)}=\Delta\mathbf x_{\rm B}^{(s)}$. Note that
$\mathcal S_{\rm B}$ depends on $\mathbf W$ and $\widehat{\mathbf x}$,
and is thus reconstructed at each candidate point. With either
construction, the continuous worst-case search over $\mathcal U_x$ is
converted into a finite-scenario problem over $\mathcal S$, which is
handled by the AO framework developed in the following subsections.

\subsection{Beamforming Optimization}
\label{subsec:w_optimization}
At the $r$-th iteration, we first update the baseband
beamforming matrices while fixing the current intended
PA-position vector $\widehat{\mathbf x}^{(r)}$.
For each $s\in\mathcal S$, the corresponding real
PA-position vector is given by
\begin{equation}
    \mathbf x_s^{(r)}
    =\widehat{\mathbf x}^{(r)}+\Delta\mathbf x^{(s)},
    \label{eq:current_scenario_position}
\end{equation}
where $\mathcal S = \mathcal S_{\rm G}$ or $\mathcal S = \mathcal S_{\rm B}$ depending on the
adopted scenario-construction strategy. At the current beamformers $\mathbf W^{(r)}$,
\Propref{Proposition_CSI} is invoked independently for every user,
subcarrier, and position-error scenario to obtain the exact
worst-case normalized CSI-error vector
$\mathbf q_{k,p}^{(s,r)}=\mathbf q_{k,p}^{\star}
\big(\mathbf W^{(r)},\mathbf x_s^{(r)}\big)$.

For a given PA deployment $\mathbf x_s^{(r)}$, define the effective channel
from the $M$ guided modes to user $k$ over subcarrier $p$ as
\begin{equation}
    \widetilde{\mathbf h}_{k,p}^{(s,r)H}
    \triangleq
    \left[
        \widehat{\mathbf h}_{k,p}
        \left(
            \mathbf x_s^{(r)}
        \right)
        +
        \epsilon_{k,p}
        \mathbf q_{k,p}^{(s,r)}
    \right]^H
    \mathbf G_p
    \left(
        \mathbf x_s^{(r)}
    \right)\in\mathbb C^{1\times M}.
    \label{eq:frozen_effective_channel}
\end{equation}

For each fixed effective channel
$\widetilde{\mathbf h}_{k,p}^{(s,r)}$, let
$u_{k,p}^{(s,r)}\in\mathbb C$ denote the receive filter.
For a candidate beamformer $\mathbf W$, the mean square error (MSE) of user $k$
over subcarrier $p$ is expressed as
\begin{equation}
\begin{aligned}
    e_{k,p}^{(s,r)}
    =&\;
    \left|u_{k,p}^{(s,r)}\right|^2
    \left(
        \sum_{j\in\mathcal K}
        \left|
            \widetilde{\mathbf h}_{k,p}^{(s,r)H}
            \mathbf w_{j,p}
        \right|^2
        +
        \sigma_{k,p}^2
    \right)
    \\
    &-
    2\operatorname{Re}
    \left\{
        u_{k,p}^{(s,r)*}
        \widetilde{\mathbf h}_{k,p}^{(s,r)H}
        \mathbf w_{k,p}
    \right\}
    +1 .
\end{aligned}
\label{eq:robust_mse}
\end{equation}
We further introduce an auxiliary MSE weight
$v_{k,p}^{(s,r)}>0$ and define the augmented weighted MSE as
\begin{equation}
    \xi_{k,p}^{(s,r)}
    \triangleq
    v_{k,p}^{(s,r)}
    e_{k,p}^{(s,r)}
    -
    \ln v_{k,p}^{(s,r)}
    -1.
    \label{eq:robust_wmse}
\end{equation}
For the current beamformers $\mathbf W^{(r)}$, minimizing
$\xi_{k,p}^{(s,r)}$ with respect to
$u_{k,p}^{(s,r)}$ and $v_{k,p}^{(s,r)}$ yields
\begin{align}
    u_{k,p}^{(s,r)}
    &=
    \frac{
        \widetilde{\mathbf h}_{k,p}^{(s,r)H}
        \mathbf w_{k,p}^{(r)}
    }{
        \displaystyle
        \sum_{j\in\mathcal K}
        \left|
            \widetilde{\mathbf h}_{k,p}^{(s,r)H}
            \mathbf w_{j,p}^{(r)}
        \right|^2
        +
        \sigma_{k,p}^2
    },
    \nonumber\\
    v_{k,p}^{(s,r)}
    &=
    1+
    \Gamma_{k,p}^{\rm wc}
    \left(
        \mathbf W_p^{(r)},
        \mathbf x_s^{(r)}
    \right),
    \label{eq:wmmse_auxiliary_updates}
\end{align}
and
\begin{equation}
    \min_{
        u_{k,p}^{(s,r)},\,
        v_{k,p}^{(s,r)}>0}
    \xi_{k,p}^{(s,r)}
    =
    -
    \ln
    \left(
        1+
        \Gamma_{k,p}^{\rm wc}
        \left(
            \mathbf W_p^{(r)},
            \mathbf x_s^{(r)}
        \right)
    \right).
    \label{eq:wmmse_equivalence}
\end{equation}

At the $r$-th iteration, the auxiliary variables
$u_{k,p}^{(s,r)}$ and $v_{k,p}^{(s,r)}$ are first updated
according to \eqref{eq:wmmse_auxiliary_updates} and then kept
fixed. To handle the worst-case position-error scenario, we
introduce an auxiliary epigraph variable $t_{\rm w}$ that upper-bounds
the aggregate WMSE for every scenario. The beamforming is optimized by solving
\begin{subequations}
\label{prob:w_update}
\begin{align}
    \min_{\mathbf W,t_{\rm w}}\quad&
    t_{\rm w}
    \label{prob:w_obj}
    \\
    \mathrm{s.t.}\quad&
    \sum_{k\in\mathcal K}
    \sum_{p\in\mathcal P}
    \xi_{k,p}^{(s,r)}
    \le t_{\rm w},
    \quad
    \forall s\in\mathcal S,
    \label{prob:w_worst_scenario}
    \\
    &
    \sum_{p\in\mathcal P}
    \xi_{k,p}^{(s,r)}
    \le c_k,
    \quad
    \forall k\in\mathcal K,\;
    \forall s\in\mathcal S,
    \label{prob:w_qos}
    \\
    &
    \|\mathbf W_p\|_F^2
    \le
    P_{p,\max},
    \quad
    \forall p\in\mathcal P,
    \label{prob:w_tone_power}
\end{align}
\end{subequations}
where $ c_k\triangleq-\frac{(P+L_{\rm CP})\ln 2}{B}R_k^{\min}$. For any given $\mathbf W$, the smallest feasible value of
$t_{\rm w}$ is $\max_{s\in\mathcal S}\sum_{k\in\mathcal K}
\sum_{p\in\mathcal P}\xi_{k,p}^{(s,r)}$. Problem~\eqref{prob:w_update} is a convex QCQP problem, which can be solved optimally by interior-point solvers, such as CVX \cite{grant2014cvx}. In what follows, a more computationally efficient
algorithm is presented to solve this problem based on the
KKT conditions.

\begin{proposition}[\textbf{KKT-Based Beamforming Update}]
\label{KKT_W}
Assume that Problem~\eqref{prob:w_update} is feasible and
satisfies Slater's condition. Let
$\theta_s\ge0, s \in \mathcal S$, $\lambda_{k,s}\ge0, k\in \mathcal K, s \in \mathcal S$ and $\rho_p\ge0, p \in \mathcal P$ denote the dual
variables associated with
\eqref{prob:w_worst_scenario},
\eqref{prob:w_qos}, and \eqref{prob:w_tone_power}, respectively. Then, an optimal solution of Problem~\eqref{prob:w_update} satisfies
\begin{equation}
    (
        \mathbf Q_p^{\rm W}
        +
        \rho_p\mathbf I_M
    )
    \mathbf W_p^\star
    =
    \mathbf Z_p^{\rm W},
    \qquad
    p\in\mathcal P,
    \label{eq:w_kkt_structure}
\end{equation}
where
\begin{equation}
\begin{aligned}
    \mathbf Q_p^{\rm W}
    \triangleq
    \sum_{s\in\mathcal S}
    \sum_{k\in\mathcal K}
    \left(
        \theta_s+\lambda_{k,s}
    \right)
    v_{k,p}^{(s,r)}
    \left|
        u_{k,p}^{(s,r)}
    \right|^2
    \widetilde{\mathbf h}_{k,p}^{(s,r)}
    \widetilde{\mathbf h}_{k,p}^{(s,r)H},
\end{aligned}
\label{eq:Qw_def}
\end{equation}
and $ \mathbf Z_p^{\rm W}
    =
    \left[
        \mathbf z_{1,p}^{\rm W},
        \ldots,
        \mathbf z_{K,p}^{\rm W}
    \right]$, with
\begin{equation}
    \mathbf z_{k,p}^{\rm W}
    \triangleq
    \sum_{s\in\mathcal S}
    \left(
        \theta_s+\lambda_{k,s}
    \right)
    v_{k,p}^{(s,r)}
    u_{k,p}^{(s,r)}
    \widetilde{\mathbf h}_{k,p}^{(s,r)}.
    \label{eq:zw_def}
\end{equation}
Moreover, the dual variables satisfy $\sum_{s\in\mathcal S}\theta_s=1$.
\end{proposition}

\emph{Proof: See Appendix~\ref{app:proof_w_update}.}

\Propref{KKT_W} provides a KKT-based solution to the convex
beamforming subproblem, where the beamforming matrices are obtained by
solving \eqref{eq:w_kkt_structure} independently over each subcarrier.
Note that Problem~\eqref{prob:w_update} is constructed with the
CSI-error vectors fixed at their worst-case values for
$\mathbf W^{(r)}$. For a different $\mathbf W$, the worst-case CSI
errors change accordingly. Hence, Problem~\eqref{prob:w_update} only
provides a local approximation of the robust objective. Therefore, the solution of Problem~\eqref{prob:w_update}, denoted by
$\mathbf W_{\rm cand}$, is used as a search direction, and the
beamformers are updated as
\begin{equation}
    \mathbf W^{(r+1)}
    =
    \mathbf W^{(r)}
    +
    \alpha_{\rm W}
    \big(\mathbf W_{\rm cand}-\mathbf W^{(r)}\big),
    \label{eq:w_line_search}
\end{equation}
where $\alpha_{\rm W}\in\{1,\frac{1}{2},\ldots,\frac{1}{16}\}$. Each candidate point is evaluated exactly via
\Propref{Proposition_CSI} over the scenarios in $\mathcal S$,
where $\mathcal S_{\rm B}$ is reconstructed at the candidate point for the
second-order construction. A candidate point is accepted according to the
following rule: If $\mathbf W^{(r)}$ satisfies
\eqref{eq:finite_position_qos}, a candidate point is accepted only if it
satisfies \eqref{eq:finite_position_qos} and does not decrease
$J_{\mathcal S}$ in \eqref{eq:finite_scenario_objective}. Otherwise, a
candidate point is accepted only if it increases the minimum worst-case user
rate $\min_{k\in\mathcal K,s\in\mathcal S}
R_k^{\rm wc}(\mathbf W,\mathbf x_s^{(r)})$. Among the accepted candidate
points, the one with the largest objective is selected. If no candidate
point is accepted, we set $\mathbf W^{(r+1)}=\mathbf W^{(r)}$.

\subsection{PA-Position Optimization}
\label{subsec:x_optimization}
We next optimize the intended PA positions while fixing the
updated beamformers $\mathbf W^{(r+1)}$. \Propref{Proposition_CSI} is first re-invoked using
$\mathbf W^{(r+1)}$ and $\mathbf x_s^{(r)}$ to obtain an
exact worst-case normalized CSI-error vector
$\mathbf q_{k,p}^{\mathrm{x},(s,r)}$ for every user,
subcarrier, and position-error scenario. These vectors are
kept fixed while constructing the local PA-position
subproblem.

Let $u_{k,p}^{\mathrm x,(s,r)}$ and $v_{k,p}^{\mathrm x,(s,r)}$ denote the corresponding MMSE receive filters and weights at the current point
$\widehat{\mathbf x}^{(r)}$. Using
\eqref{eq:robust_mse}--\eqref{eq:robust_wmse}, the
resulting weighted MSE can be denoted as
$\xi_{k,p}^{\mathrm{x},(s,r)}$.
The PA-position subproblem can then be written as
\begin{subequations}
\label{prob:x_original}
\begin{align}
    \min_{\widehat{\mathbf x},t_{\rm x}}\quad&
    t_{\rm x}
    \label{prob:x_original_obj}\\
    \mathrm{s.t.}\quad&
    \sum_{k\in\mathcal K}
    \sum_{p\in\mathcal P}
    \xi_{k,p}^{\mathrm{x},(s,r)}
    \le t_{\rm x},
    \quad
    \forall s\in\mathcal S,
    \label{prob:x_original_worst}\\
    &
    \sum_{p\in\mathcal P}
    \xi_{k,p}^{\mathrm{x},(s,r)}
    \le
    c_k,
    \forall k\in\mathcal K,\;
    \forall s\in\mathcal S,
    \label{prob:x_original_qos}\\
 &\eqref{prob:position_boundary}, ~
    \eqref{prob:position_spacing}.
    \label{prob:x_original_spacing}
\end{align}
\end{subequations}
where $t_{\rm x}$ is an auxiliary epigraph variable. Problem~\eqref{prob:x_original} remains non-convex since
both $\widehat{\mathbf h}_{k,p}(\mathbf x_s)$ and
$\mathbf G_p(\mathbf x_s)$ depend nonlinearly on the PA
positions. We therefore construct a convex local surrogate
using SCA.

At the $r$-th iteration, given
$\widehat{\mathbf x}^{(r)}$ and the updated beamformers
$\mathbf W^{(r+1)}$, we construct the following convex
quadratic surrogate:
{
\begin{equation}
\begin{aligned}
   & \widetilde{\xi}_{k,p}^{\mathrm x,(s,r)}
    \left(
        \widehat{\mathbf x};
        \widehat{\mathbf x}^{(r)}
    \right)\\
   & \triangleq
    \xi_{k,p}^{\mathrm x,(s,r)}
    +
    (\nabla_{\widehat{\mathbf x}}
    \xi_{k,p}^{\mathrm x,(s,r)})^T
    \left(
        \widehat{\mathbf x}
        -
        \widehat{\mathbf x}^{(r)}
    \right)
    \\
    &+
    \frac{1}{2}
    \left(
        \widehat{\mathbf x}
        -
        \widehat{\mathbf x}^{(r)}
    \right)^T
   \left( \mathbf T_{k,p}^{(s,r)} +
        \iota^{(r)}\mathbf I_N \right)
    \left(
        \widehat{\mathbf x}
        -
        \widehat{\mathbf x}^{(r)}
    \right),
\end{aligned}
\label{eq:x_wmse_surrogate}
\end{equation}}
\hspace*{-\parindent}~where
$\mathbf T_{k,p}^{(s,r)}
\in\mathbb R^{N\times N}$ and
$\mathbf T_{k,p}^{(s,r)}\succeq\mathbf 0$
denotes a Gauss--Newton curvature matrix and $\iota^{(r)}>0$ is a damping
parameter.

To derive the gradient in
\eqref{eq:x_wmse_surrogate} for
$k,j\in\mathcal K$, we first define
\begin{equation}
\begin{aligned}
    \omega_{k,j,p}^{(s,r)}
    \left(
        \widehat{\mathbf x}
    \right)
    \triangleq\;&
    \left[
        \widehat{\mathbf h}_{k,p}
        \left(
            \widehat{\mathbf x}
            +
            \Delta\mathbf x^{(s)}
        \right)
        +
        \epsilon_{k,p}
        \mathbf q_{k,p}^{\mathrm x,(s,r)}
    \right]^H
    \\
    &\times
    \mathbf G_p
    \left(
        \widehat{\mathbf x}
        +
        \Delta\mathbf x^{(s)}
    \right)
    \mathbf w_{j,p}^{(r+1)} .
\end{aligned}
\label{eq:x_t_def}
\end{equation}
Since $\Delta\mathbf x^{(s)}$ is known for each scenario,
the derivative of $\omega_{k,j,p}^{(s,r)}\left(\widehat{\mathbf x}\right)$ with respect to $\widehat x_n$ is identical
to that with respect to the corresponding real position $x_n$. Therefore, we can obtain
\par{\small
\begin{equation}
\begin{aligned}
    \left.
    \frac{
        \partial
        \omega_{k,j,p}^{(s,r)}(\widehat{\mathbf x})
    }{
        \partial \widehat x_n
    }
    \right|_{\widehat{\mathbf x}
    =\widehat{\mathbf x}^{(r)}}
    \!\!\!\!\!\!\!=&\;
    \Bigg[
        \mathbf J_{h,k,p,n}^{(s,r)H}
        \mathbf G_p
        \left(
            \mathbf x_s^{(r)}
        \right)
    \\
    +& \left(
            \widehat{\mathbf h}_{k,p}
            \left(
                \mathbf x_s^{(r)}
            \right)
            +
            \epsilon_{k,p}
            \mathbf q_{k,p}^{\mathrm x,(s,r)}
        \right)^H
        \mathbf J_{G,p,n}^{(s,r)}
    \Bigg]
    \mathbf w_{j,p}^{(r+1)},
\end{aligned}
\label{eq:x_t_derivative}
\end{equation}}
\hspace*{-\parindent}~where
$\mathbf J_{h,k,p,n}^{(s,r)}$ and
$\mathbf J_{G,p,n}^{(s,r)}$ denote the derivatives of
$\widehat{\mathbf h}_{k,p}(\mathbf x)$ and
$\mathbf G_p(\mathbf x)$ with respect to $x_n$, respectively,
evaluated at $\mathbf x_s^{(r)}$.

The gradient of the weighted MSE in \eqref{eq:x_wmse_surrogate} is given as
\begin{equation}
\begin{aligned}
    \nabla_{\widehat{\mathbf x}}
    \xi_{k,p}^{\mathrm x,(s,r)}
    =&\;
    2v_{k,p}^{\mathrm x,(s,r)}
    |u_{k,p}^{\mathrm x,(s,r)}|^2
    {\rm Re}
    \left\{
        \sum_{j\in\mathcal K}
        \omega_{k,j,p}^{(s,r)}
        \left(
            \widehat{\mathbf x}^{(r)}
        \right)^{*}
        \mathbf f_{k,j,p}^{(s,r)}
    \right\} 
    \\
    &-
    2v_{k,p}^{\mathrm x,(s,r)}
    {\rm Re}
    \left\{
        u_{k,p}^{\mathrm x,(s,r)*}
        \mathbf f_{k,k,p}^{(s,r)}
    \right\},
\end{aligned}
\label{eq:x_wmse_gradient}
\end{equation}
where $ \mathbf f_{k,j,p}^{(s,r)} \triangleq \big[\nabla_{\widehat x_1}\omega_{k,j,p} \big|_{\widehat{\mathbf x}
    =\widehat{\mathbf x}^{(r)}}, \cdots, \nabla_{\widehat x_N}\omega_{k,j,p} \big|_{\widehat{\mathbf x}
    =\widehat{\mathbf x}^{(r)}}\big]^T$.

The Gauss--Newton curvature matrix in
\eqref{eq:x_wmse_surrogate} is constructed as
\begin{equation}
    \mathbf T_{k,p}^{(s,r)}
    =
    2v_{k,p}^{\mathrm x,(s,r)}
    \left|u_{k,p}^{\mathrm x,(s,r)}\right|^2
    \sum_{j\in\mathcal K}
    {\rm Re}
    \left\{
        \mathbf f_{k,j,p}^{(s,r)}
        \mathbf f_{k,j,p}^{(s,r)H}
    \right\}.
    \label{eq:x_gn_curvature}
\end{equation}
The matrix $\mathbf T_{k,p}^{(s,r)}$ provides a positive
semidefinite Gauss--Newton approximation of the local weighted-MSE
curvature. The damping parameter is adaptively chosen as
\begin{equation}
    \iota^{(r)}=
    \frac{\tilde \iota}{|\mathcal S| KPN}
    {\rm tr}
    \left(
        \sum_{s\in\mathcal S}
        \sum_{k\in\mathcal K}
        \sum_{p\in\mathcal P}
        \mathbf T_{k,p}^{(s,r)}
    \right)
    +
    \iota_{\min},
    \iota_{\min}>0,
    \label{eq:x_damping}
\end{equation}
where $\tilde \iota>0$ is a scaling factor.

By replacing the weighted-MSE terms with the surrogate
functions in \eqref{eq:x_wmse_surrogate}, the candidate
intended PA-position vector is obtained by solving
\begin{subequations}
\label{prob:x_update}
\begin{align}
    \min_{\widehat{\mathbf x},\,t_{\rm x}}
    &~
    t_{\rm x}
    \label{prob:x_obj}
    \\
    \mathrm{s.t.}&
    \sum_{k\in\mathcal K}
    \sum_{p\in\mathcal P}
    \widetilde{\xi}_{k,p}^{\mathrm x,(s,r)}
    \left(
        \widehat{\mathbf x};
        \widehat{\mathbf x}^{(r)}
    \right)
    \le
    t_{\rm x},
    \forall s\in\mathcal S,
    \label{prob:x_worst_scenario}
    \\
    &
    \sum_{p\in\mathcal P}
    \widetilde{\xi}_{k,p}^{\mathrm x,(s,r)}
    \left(
        \widehat{\mathbf x};
        \widehat{\mathbf x}^{(r)}
    \right)
    \le
    c_k,
    \forall k\in\mathcal K,\;
    \forall s\in\mathcal S,
    \label{prob:x_qos}
    \\
 &\eqref{prob:position_boundary}, ~
    \eqref{prob:position_spacing}.
    \label{prob:x_spacing}
\end{align}
\end{subequations}
Since $\mathbf T_{k,p}^{(s,r)}+\iota^{(r)}\mathbf I_N
\succ\mathbf 0$,
the surrogate functions in
\eqref{eq:x_wmse_surrogate} are strongly convex with respect
to $\widehat{\mathbf x}$. Hence,
Problem~\eqref{prob:x_update} is a convex QCQP. Let $\widehat{\mathbf x}_{\rm cand}^{(r+1)}$ denote the solution of Problem~\eqref{prob:x_update}. The intended PA-position vector is then updated as
\begin{equation}
    \widehat{\mathbf x}^{(r+1)}
    =
    \widehat{\mathbf x}^{(r)}
    +
    \alpha_{\rm x}^{(r+1)}
    \left(
        \widehat{\mathbf x}_{\rm cand}^{(r+1)}
        -
        \widehat{\mathbf x}^{(r)}
    \right),
    \label{eq:x_sca_update}
\end{equation}
where $\alpha_{\rm x}^{(r+1)}\in(0,1]$ denotes the SCA step size. The step size is determined by a backtracking line
search~\cite{nocedal2006numerical} under the same acceptance rule as
the beamforming update. If no step size is accepted, we set
$\widehat{\mathbf x}^{(r+1)}=\widehat{\mathbf x}^{(r)}$. Moreover, the PA positions are updated only if \eqref{eq:finite_position_qos} is satisfied.

\begin{proposition}[\textbf{KKT-Based PA-Position Update}]
\label{Proposition_x}
Assume that Problem~\eqref{prob:x_update} is feasible and
satisfies Slater's condition. Let $\chi_s\ge0$ and
$\eta_{k,s}\ge0$ denote the dual variables associated with
\eqref{prob:x_worst_scenario} and \eqref{prob:x_qos},
respectively. Furthermore, let
$\varrho_n^{\mathrm{L}}\ge0$,
$\varrho_n^{\mathrm{U}}\ge0$, and $\varphi_n\ge0$
denote the dual variables associated with the lower
PA-position bound, the upper PA-position bound, and
the spacing constraint between PAs $n$ and $n+1$,
respectively. Then, $\widehat{\mathbf x}_{\rm cand}^{(r+1)}$ satisfies
\begin{equation}
\label{eq:x_kkt_stationarity}
\begin{aligned}
    &\mathbf \Xi^{\mathrm x}
    \left(
        \widehat{\mathbf x}_{\rm cand}^{(r+1)}
        -
        \widehat{\mathbf x}^{(r)}
    \right)
    +
    \boldsymbol\Lambda^{\mathrm x}
    \\
    &
    -
    \sum_{n=1}^{N}
   \varrho_n^{\mathrm{L}}\mathbf e_n
    +
    \sum_{n=1}^{N}
    \varrho_n^{\mathrm{U}}\mathbf e_n
    +
    \sum_{n=1}^{N-1}
    \varphi_n
    \left(
        \mathbf e_n-\mathbf e_{n+1}
    \right)
    =
    \mathbf 0,
\end{aligned}
\end{equation}
where $\mathbf e_n\in\mathbb R^{N}$ denotes the $n$th
canonical basis vector,
\begin{align}
    \mathbf \Xi^{\mathrm x}
    &\triangleq
    \sum_{s\in\mathcal S}
    \sum_{k\in\mathcal K}
    (\chi_s+\eta_{k,s})
    \sum_{p\in\mathcal P}
    \big(\mathbf T_{k,p}^{(s,r)}+\iota^{(r)}\mathbf I_N\big),
    \label{eq:x_Xi_def}\\
    \boldsymbol\Lambda^{\mathrm x}
    &\triangleq
    \sum_{s\in\mathcal S}
    \sum_{k\in\mathcal K}
    (\chi_s+\eta_{k,s})
    \sum_{p\in\mathcal P}
    \nabla_{\widehat{\mathbf x}}
    \xi_{k,p}^{\mathrm x,(s,r)}.
    \label{eq:x_zeta_def}
\end{align}
Moreover, the dual variables satisfy
$\sum_{s\in\mathcal S}\chi_s=1$, and
$\mathbf\Xi^{\mathrm x}\succ\mathbf 0$. Hence,
\eqref{eq:x_kkt_stationarity} can be equivalently written as
\begin{equation}
\begin{aligned}
\widehat{\mathbf x}_{\rm cand}^{(r+1)}
=&\widehat{\mathbf x}^{(r)}
-\big(\mathbf\Xi^{\mathrm x}\big)^{-1}
\Big(\boldsymbol\Lambda^{\mathrm x}
-\sum_{n=1}^{N}\big(\varrho_n^{\mathrm L}-\varrho_n^{\mathrm U}\big)\mathbf e_n\\
&+\sum_{n=1}^{N-1}\varphi_n\big(\mathbf e_n-\mathbf e_{n+1}\big)\Big).
\end{aligned}
\label{eq:x_closed_form}
\end{equation}
\end{proposition}

\emph{Proof: See Appendix~\ref{app:proof_x_update}.}

\Propref{Proposition_x} shows that the candidate PA-position vector is obtained in closed form via
\eqref{eq:x_closed_form} for given dual variables. The KKT conditions, including the dual
variables, are solved by an active-set Newton method. Since the SCA model locally approximates the nonlinear position-dependent channels, the resulting candidate is used to update the intended PA positions through \eqref{eq:x_sca_update} under the acceptance rule.

\subsection{Multi-Start Initialization}
\label{subsec:initialization}

Since the joint beamforming and PA-position problem is
non-convex, the AO algorithm can be sensitive to the initial
PA deployment. We therefore employ a robust-aware multi-start
initialization. Specifically, a candidate pool
$\mathcal C_{\rm x}$ is first generated, where each
$\widehat{\mathbf x}^{(c)}\in\mathcal C_{\rm x}$ satisfies
the robust PA-deployment constraints in
Problem~\eqref{prob:robust_design}.

For each candidate, define the effective channel based on the
estimated CSI at the intended PA positions as
\begin{equation}
    \mathbf H^{\rm e}_p=[\mathbf G_p^H\widehat{\mathbf h}_{1,p},\ldots,\mathbf G_p^H\widehat{\mathbf h}_{K,p}]^H\in\mathbb C^{K\times M},
\end{equation}
where all position-dependent quantities are evaluated at
$\widehat{\mathbf x}^{(c)}$. The corresponding RZF
beamformer is given by
\begin{equation}
\mathbf W_{p,\rm RZF}^{(c)}
\propto
\mathbf H_p^{\rm e,H}
\Big(
\mathbf H_p^{\rm e}\mathbf H_p^{\rm e,H}
+\frac{KP \bar \sigma^2}{P_{\max}}\mathbf I_K
\Big)^{-1},
\quad p\in\mathcal P,
\label{eq:init_rzf}
\end{equation}
which is normalized such that
$\|\mathbf W_{p,\rm RZF}^{(c)}\|_F^2=P_{\max}/P$, where
$\bar\sigma^2\triangleq\frac{1}{KP}\sum_{k\in\mathcal K}
\sum_{p\in\mathcal P}\sigma_{k,p}^2$ denotes the average noise power.

To reduce the initialization complexity, the candidate pool is
processed in two stages. First, all candidates are evaluated
using the RZF beamformers in \eqref{eq:init_rzf} under the
estimated channels at the intended PA positions, i.e., with
zero CSI and PA positioning errors, and only the highest-rate
candidates are retained. The reduced set is then re-evaluated
using the nonlinear channel model over all sampled
position-error scenarios, while \Propref{Proposition_CSI} is invoked to
obtain the exact worst-case CSI performance for each user and
subcarrier.

The retained candidates are ranked according to
$J_{\mathcal S}$ in \eqref{eq:finite_scenario_objective},
while their finite-scenario QoS feasibility is checked
separately.
The $L_{\rm ini}$ highest-ranked nonduplicate candidates are retained as the AO starting points. All retained starts enter the robust AO iterations, where a start violating \eqref{eq:finite_position_qos} is first driven toward QoS feasibility by the acceptance rule in Section~\ref{subsec:w_optimization}. After all retained starts have been optimized
independently, the feasible solution achieving the largest
finite-scenario robust objective is selected as the final
solution.

\subsection{Convergence and Complexity Analysis}
Due to the acceptance rule, the minimum worst-case user rate
increases at each accepted update until the QoS constraints are
satisfied, after which the QoS constraints are preserved and
$J_{\mathcal S}$ in \eqref{eq:finite_scenario_objective} is
non-decreasing. For the second-order construction, each point is
evaluated with its own scenario set. Since $\mathcal S_{\rm B}$ is
uniquely determined by $(\mathbf W,\widehat{\mathbf x})$, the resulting
objective is a well-defined function of the iterate, to which the
acceptance rule applies.
Since the achievable sum rate is upper-bounded under the
transmit-power constraints, the sequence of objective values converges.
The resulting procedure is summarized in
\textbf{Algorithm~\ref{alg:robust_ao}}. For a given scenario set
$\mathcal S\in\{\mathcal S_{\rm G},\mathcal S_{\rm B}\}$, the dominant computational complexity
of the beamforming and PA-position updates is
\par{\small
\begin{equation}
\begin{aligned}
\mathcal O\Big(
I_{\rm W}\big(|\mathcal S|KPM^2+PM^3\big)+|\mathcal S|PK^2N(M+N)+I_{\rm x}N^3
\Big),
\end{aligned}
\label{eq:overall_ao_complexity}
\end{equation}}
\hspace*{-\parindent}~\,where $I_{\rm W}$ and $I_{\rm x}$ denote the numbers of
inner KKT iterations for the beamforming and PA-position
updates, respectively. The first two terms account for constructing
$\mathbf Q_p^{\rm W}$ and $\mathbf Z_p^{\rm W}$ and
solving the $M\times M$ linear systems, while the
remaining terms correspond to the PA-position gradient,
curvature, and KKT computations. For the grid-based construction,
$|\mathcal S|=|\mathcal S_{\rm G}|=Q^N$. For the second-order
construction, solving the $K+1$ quadratic problems requires
enumerating $3^N$ active-bound patterns, which only involves algebraic
operations, and $|\mathcal S_{\rm B}|$ equals the number of retained
local minimizers. Moreover, each exact evaluation of
$J_{\mathcal S}$ via \Propref{Proposition_CSI} requires
solving $|\mathcal S|KP$ low-dimensional trust-region subproblems.

\begin{algorithm}[t]
\caption{Robust Joint Beamforming and PA-Position Optimization}
\label{alg:robust_ao}
\renewcommand{\algorithmicrequire}{\textbf{Input:}}
\renewcommand{\algorithmicensure}{\textbf{Output:}}
\begin{algorithmic}[1]
\REQUIRE
$\{P_{p,\max}\}_{p\in\mathcal P}$, $\{R_k^{\min}\}_{k\in\mathcal K}$,
$\mathcal U_x$, the adopted scenario-construction strategy,
$L_{\rm ini}$, and $\varepsilon_{\rm AO}$.
\ENSURE
$\mathbf W^{\star}$ and $\widehat{\mathbf x}^{\star}$.
\STATE Generate $L_{\rm ini}$ initial points
$\{(\mathbf W_{\ell}^{(0)},\widehat{\mathbf x}_{\ell}^{(0)})\}_{\ell=1}^{L_{\rm ini}}$
via the multi-start initialization in
Section~\ref{subsec:initialization}.
\FOR{$\ell=1,\ldots,L_{\rm ini}$}
    \STATE Set $r=0$ and $(\mathbf W^{(0)},\widehat{\mathbf x}^{(0)})
    =(\mathbf W_{\ell}^{(0)},\widehat{\mathbf x}_{\ell}^{(0)})$.
    \REPEAT
        \STATE Construct $\mathcal S$
        and $\{\mathbf x_s^{(r)}\}_{s\in\mathcal S}$ according to
        Section~\ref{subsec:position_scenario_construction}, obtain
        $\mathbf q_{k,p}^{(s,r)}$ via \Propref{Proposition_CSI},
        update the WMMSE variables via
        \eqref{eq:wmmse_auxiliary_updates}, and update
        $\mathbf W^{(r+1)}$ via \Propref{KKT_W} and
        \eqref{eq:w_line_search} under the acceptance rule.
        \STATE Reconstruct $\mathcal S$
        at $(\mathbf W^{(r+1)},\widehat{\mathbf x}^{(r)})$, obtain
        $\mathbf q_{k,p}^{\mathrm x,(s,r)}$ via
        \Propref{Proposition_CSI}, and update
        $\widehat{\mathbf x}^{(r+1)}$ via \Propref{Proposition_x} and
        \eqref{eq:x_sca_update} under the acceptance rule.
        \STATE Set $r=r+1$.
    \UNTIL{$|J_{\mathcal S}(\mathbf W^{(r)},\widehat{\mathbf x}^{(r)})
    -J_{\mathcal S}(\mathbf W^{(r-1)},\widehat{\mathbf x}^{(r-1)})|
    \le\varepsilon_{\rm AO}
    J_{\mathcal S}(\mathbf W^{(r-1)},\widehat{\mathbf x}^{(r-1)})$
    and \eqref{eq:finite_position_qos} is satisfied, or the maximum
    number of iterations is reached}
\ENDFOR
\STATE Select the feasible start with the largest $J_{\mathcal S}$ as
$(\mathbf W^{\star},\widehat{\mathbf x}^{\star})$.
\end{algorithmic}
\end{algorithm}

\section{Numerical Results}
\label{sec:numerical}

\begin{table}[t]
\caption{Simulation Parameters}
\label{tab:simulation_parameters}
\centering
\footnotesize
\setlength{\tabcolsep}{3pt}
\renewcommand{\arraystretch}{1.12}
\begin{tabular}{@{}p{0.62\columnwidth}p{0.32\columnwidth}@{}}
\hline
\textbf{Parameter} & \textbf{Value} \\
\hline
Numbers of users, modes, and PAs $(K,M,N)$ & $(3,3,3)$ \\
Carrier frequency $f_{\mathrm c}$ & $28$ GHz \\
System bandwidth $B$ & $500$ MHz \\
Number of subcarriers $P$ & $256$ \\
CP length $L_{\mathrm{CP}}$ & $128$ \\
Total transmit power $P_{\max}$ & $0.1$ W \\
Waveguide relative permittivity & $4$ \\
Waveguide width and height
    & $8$ mm, $7$ mm \\
Waveguide propagation loss at $f_{\mathrm c}$
    & $0.01$ dB/m \\
Waveguide length $L_{\mathrm{wg}}$ & $20$ m \\
PA deployment height $h_{\mathrm{PA}}$ & $2.5$ m \\
Noise power spectral density
    & $-174$ dBm/Hz \\
Receiver noise figure & $0$ dB \\
Minimum user rate $R_k^{\min}, k \in \mathcal K$
    & $0.5$ Gbps \\
Coupling coefficients $(\kappa_{\mathrm c},\kappa_{\rm f})$
    & $(10,5)$ m$^{-1}$ \\
Mode-leakage factor $\mu_{\mathrm{leak}}$
    & $0.3$ \\
Inter-mode coupling factor $\eta_{\mathrm{ww}}$
    & $0.02$ \\
CSI error radius $\epsilon_{k,p}$
    & $6\times10^{-6}$ \\
PA positioning error bound $\delta_x$
    & $0.04\lambda_c$ \\
PA deployment region $[x_{\min},x_{\max}]$
    & $[0.0785,\,19.9215]$ m \\
Minimum PA spacing $\Delta_{\min}$
    & $0.1571$ m \\
Number of AO starting points $L_{\rm ini}$
    & $6$ \\
AO convergence tolerance $\varepsilon_{\rm AO}$
    & $10^{-4}$ \\
\hline
\end{tabular}
\end{table}

To evaluate the effectiveness of the proposed robust designs, we
consider a downlink MPA-OFDM system, where an open dielectric
rectangular waveguide extends along the $x$-axis at a height of
$h_{\mathrm{PA}}$ above the user plane. User $k$ is located at
$(a_k,b_k,0)$, where $a_k$ and $b_k$ are independently and uniformly
distributed over $[3,20]$ m and $[3,10]$ m, respectively. The effective
indices of the three guided modes are determined using the
effective-index method (EIM) for rectangular dielectric
waveguides~\cite{okamoto2006fundamentals}, which yields
$\mathbf n_{\mathrm{eff}}(f_{\mathrm c})=[1.83496,1.59977,1.17512]$ at
the carrier frequency. To incorporate frequency dispersion, the
propagation constant of guided mode $m$ at subcarrier $p$ is calculated
as
\begin{equation}
\beta_{m,p}
=\frac{2\pi f_{\mathrm c}}{c}
n_{\mathrm{eff},m}(f_{\mathrm c})
+\beta_m^{\mathrm{EIM}}(f_p)
-\beta_m^{\mathrm{EIM}}(f_{\mathrm c}),
\label{eq:simulation_dispersion}
\end{equation}
where $\beta_m^{\mathrm{EIM}}(f)$ is the frequency-dependent propagation
constant predicted by the EIM.

We set $M=N=3$ and adopt a one-to-one PA-to-mode assignment is adopted, with PA
$n$ associated with guided mode $m=n$. The propagation index of each PA
is chosen as $n_{\mathrm{PA},n}=n_{\mathrm{eff},n}(f_{\mathrm c})$, and
its propagation constant is given by
$\beta_{\mathrm{PA},n,p}=2\pi f_p n_{\mathrm{PA},n}/c$, which ensures
phase matching between the assigned PA and guided mode at
$f_{\mathrm c}$. Following~\cite{xiao2025frequency}, the
frequency-dependent PA-mode coupling coefficient is modeled as
$\kappa_p=\kappa_{\mathrm c}+\kappa_{\rm f}(f_p-f_{\mathrm c})/f_{\mathrm c}$.
For PA $n$, the coupling coefficient associated with guided mode $m'$ is
set to $\kappa_{n,m',p}=\kappa_p$ if $m'=n$, and
$\kappa_{n,m',p}=\mu_{\mathrm{leak}}\kappa_p$ otherwise. Moreover, weak
direct coupling between distinct guided modes within each PA coupling
region is incorporated through
$\kappa_{n,m,m',p}^{\mathrm{ww}}=\eta_{\mathrm{ww}}\kappa_p$ for
$m\neq m'$, and the coupling length of each PA is set to
$L_n=\pi/(2\kappa_{\mathrm c})$. Unless otherwise stated, the CSI error
radius is identical for all users and subcarriers, i.e.,
$\epsilon_h=\epsilon_{k,p}=6\times10^{-6}$, and the normalized PA
positioning error radius, defined as
$\upsilon_x\triangleq\delta_x/\lambda_c$ with $\lambda_c=c/f_{\mathrm c}$
being the free-space wavelength, is set to $\upsilon_x=0.04$. The other
default simulation parameters are summarized in
Table~\ref{tab:simulation_parameters}.

The following schemes are considered for comparison.

\begin{itemize}

\item \textbf{Proposed-G (Grid-Based):}
This scheme employs the proposed robust AO algorithm, where the worst-case
PA positioning errors are handled through grid-based
scenario construction. Specifically, $Q=2$ grid
points are used for each PA.

\item \textbf{Proposed-SO (Second-Order):}
This scheme adopts the same proposed robust AO
algorithm but employs
second-order approximation-based scenario
construction.

\item \textbf{NR (Non-Robust), PO-R
(Position-Only Robust), and CSI-R (CSI-Only Robust):}
These schemes neglect both errors,
CSI error only, and PA positioning error
only, respectively. 

\item \textbf{D-WMMSE-SCA
(Direct WMMSE-SCA):}
This benchmark extends the WMMSE-based joint design in~\cite{xu2026attenuation} to the considered robust model. It is initialized from a single deterministic PA configuration and its RZF beamformers, without robust candidate screening or multi-start.

\item \textbf{RZF-Sel
(Regularized Zero-Forcing Selection):}
This benchmark adopts RZF beamforming
\cite{bereyhi2026mimo} and selects the PA configuration
using worst-case rate and QoS evaluation, without subsequent joint
AO refinement.

\item \textbf{SM-PASS (Single-Mode Pinching-Antenna System):}
This benchmark adopts a single-mode PASS with
OFDMA \cite{oikonomou2025ofdma}.
Only the fundamental guided mode is employed,
and each subcarrier is assigned to one user.
PA positions are determined through robust
candidate screening.

\item \textbf{FO-Pos (First-Order Position Approximation):}
Motivated by the first-order position-error modeling in PA
systems~\cite{jiang2026impact}, this benchmark  employs first-order Taylor expansion with respect to PA positioning errors  to
construct approximate worst-case sum-rate and QoS metrics, based on which the beamformers and intended PA positions are jointly optimized.
\end{itemize}

For a fair comparison, all schemes are evaluated under the same CSI and PA positioning error settings. For each point, we generate 30 Monte-Carlo
simulations and obtain their average results.

\begin{figure}[t]
    \centering
    \includegraphics[scale=0.32]{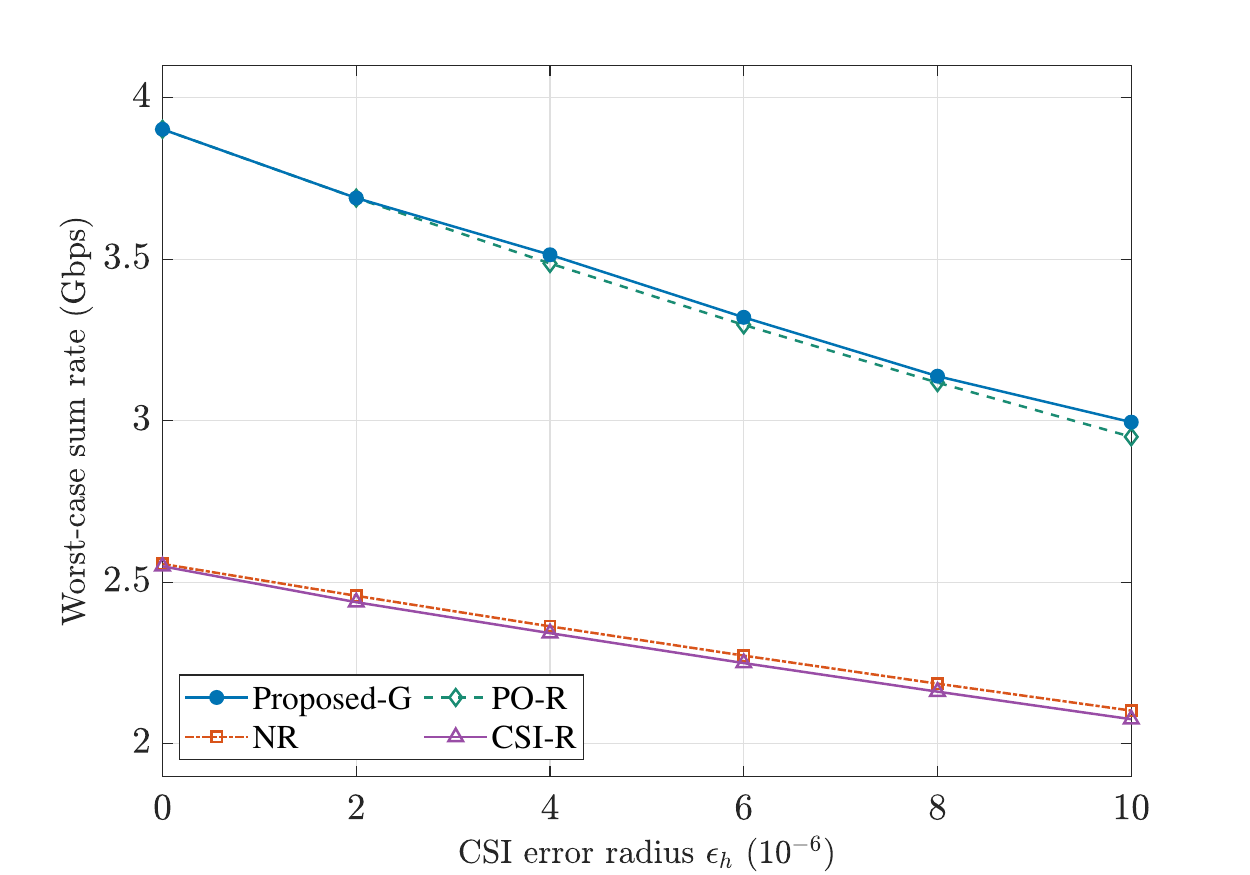}
    \caption{Worst-case sum rate versus the CSI error radius for different error modeling schemes ($\upsilon_x=0.04$).}
    \label{fig:csi_robustness}
    \vspace{-5mm}
\end{figure}

\begin{figure}[t]
    \centering
    \includegraphics[scale=0.32]{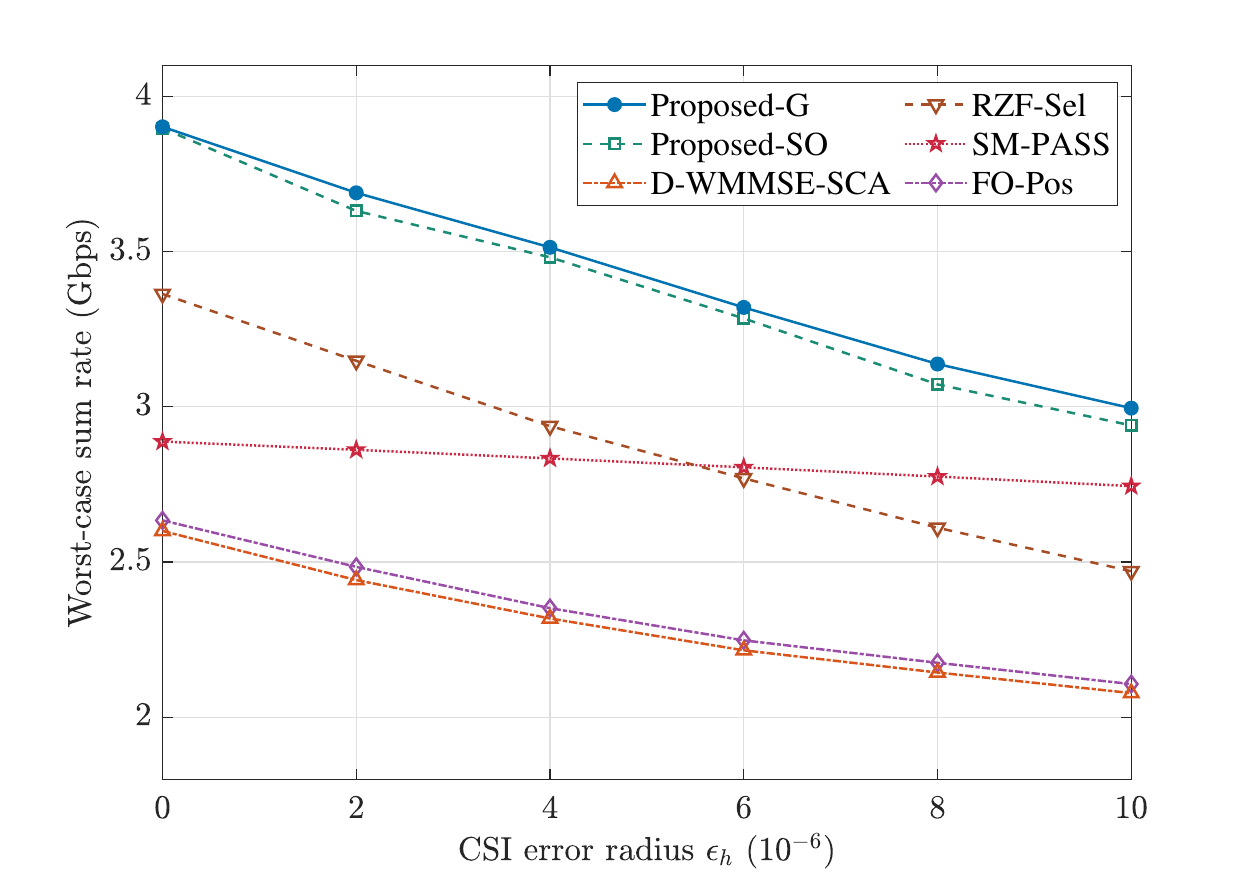}
    \caption{Worst-case sum rate versus the CSI error radius for the proposed algorithms and benchmark schemes ($\upsilon_x=0.04$).}
    \label{fig:csi_algorithms}
    \vspace{-5mm}
\end{figure}

\begin{figure}[t]
    \centering
    \includegraphics[scale=0.32]{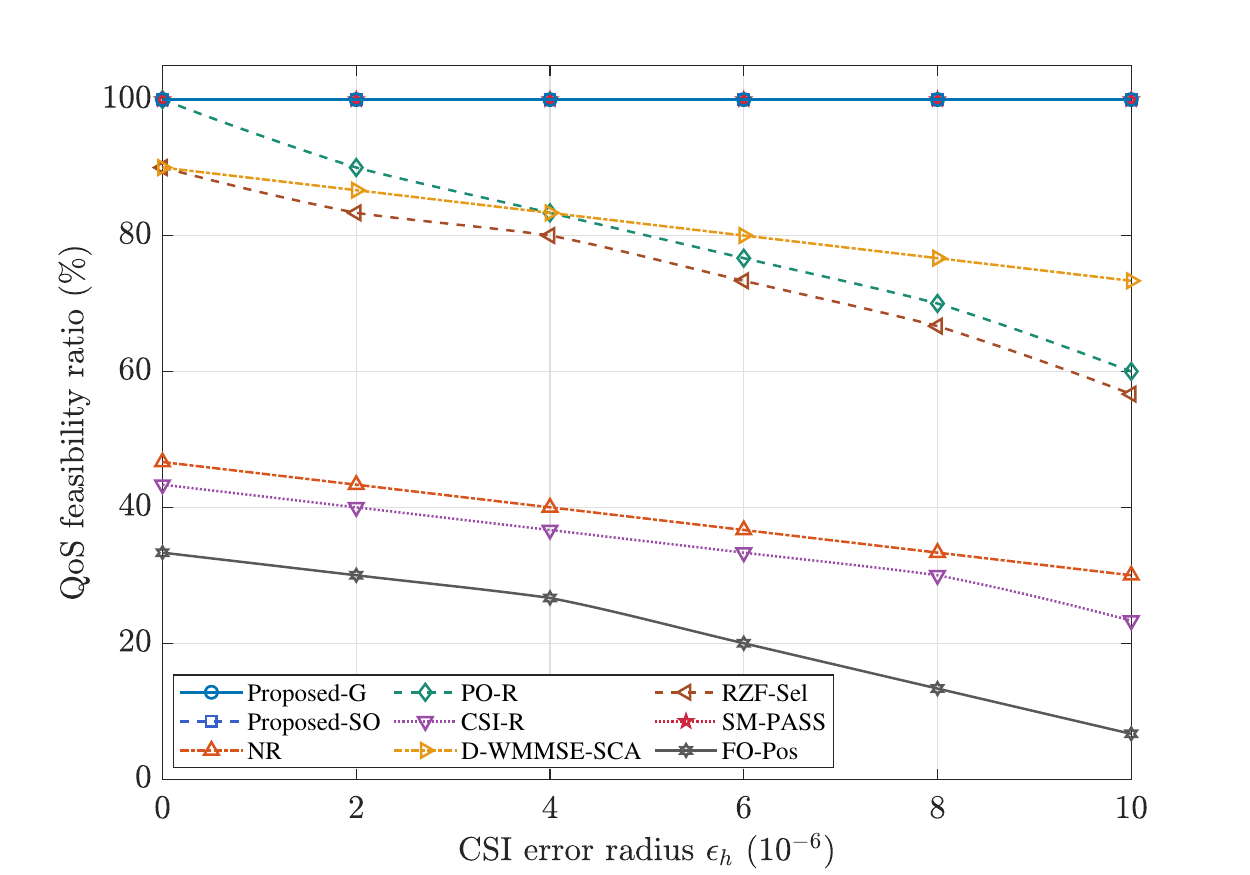}
    \caption{Worst-case QoS feasibility ratio versus the CSI error radius for different schemes ($\upsilon_x=0.04$).}
    \label{fig:csi_qos}
    \vspace{-5mm}
\end{figure}

Figure~\ref{fig:csi_robustness} shows the
worst-case sum rate versus the CSI error
radius for different error modeling schemes.
As the CSI error increases, all schemes
experience performance degradation. At
$\epsilon_h=6\times10^{-6}$, Proposed-G achieves
sum-rate gains of approximately $46.1\%$ and
$47.6\%$ over NR and CSI-R, respectively.
These gains demonstrate the importance of
accounting for PA positioning errors, which alter
the effective user channels and may lead to
reduced desired-signal power and increased
inter-user interference. Moreover,  although PO-R incurs only a $0.7\%$ sum-rate loss compared with Proposed-G, neglecting CSI errors
during optimization can lead to violations of
the user QoS constraints, as will be shown
in Fig.~\ref{fig:csi_qos}.

Figure~\ref{fig:csi_algorithms} compares the
worst-case sum rates of different schemes versus
the CSI error radius. Both proposed schemes
achieve superior performance over the considered
benchmarks. In particular, at $\epsilon_h=6\times10^{-6}$, Proposed-SO retains
approximately $98.4\%$ of the sum rate of Proposed-G, while Proposed-G
achieves sum-rate gains of approximately
$49.9\%$, $19.9\%$, and $18.4\%$ over
D-WMMSE-SCA, RZF-Sel, and SM-PASS, respectively.
These gains demonstrate the benefits of robust
initial-point selection, iterative joint optimization,
and the additional transmission flexibility
provided by multi-mode operation. Meanwhile, SM-PASS exhibits a relatively small
rate reduction as the CSI error increases, since
its OFDMA architecture avoids inter-user
interference on each subcarrier and thus reduces
the sensitivity to CSI-induced interference
variations. In contrast, FO-Pos exhibits a more pronounced
performance loss, suggesting that the first-order
approximation may not adequately capture the
nonlinear effects of PA positioning errors on
the effective user channels.

Figure~\ref{fig:csi_qos} illustrates the worst-case
QoS feasibility ratio versus the CSI error radius,
defined as the percentage of user deployments
satisfying the minimum rate requirements for all
users under the considered errors.
Proposed-G and Proposed-SO maintain $100\%$ feasibility across
all CSI error levels. Although PO-R achieves a
comparable sum rate in Fig.~\ref{fig:csi_robustness},
its QoS feasibility ratio decreases to $77\%$
and $60\%$ at $\epsilon_h=6\times10^{-6}$
and $10^{-5}$, respectively. This indicates that
ignoring CSI errors during optimization may cause some users to fail to meet their requirements despite achieving a high sum rate. Meanwhile, SM-PASS also maintains $100\%$ feasibility under its single-mode OFDMA architecture.
In contrast, D-WMMSE-SCA and RZF-Sel exhibit
lower feasibility ratios, whereas NR, CSI-R,
and FO-Pos experience more frequent QoS violations.
These results underscore the importance of
jointly handling both errors to ensure reliable QoS.

\begin{figure}[t]
    \centering
    \includegraphics[scale=0.32]{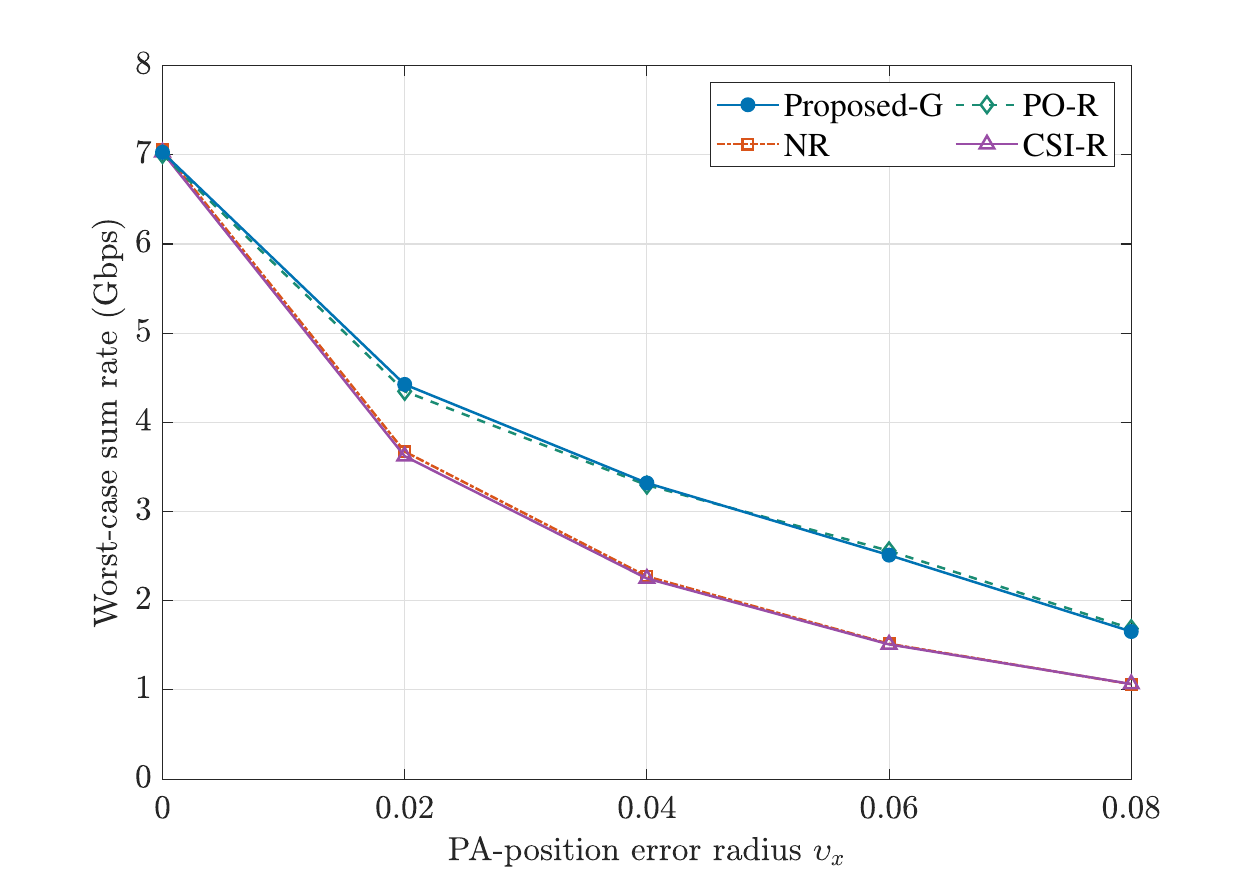}
    \caption{Worst-case sum rate versus the normalized
    PA positioning error radius for different error
    modeling schemes ($\epsilon_h=6\times10^{-6}$).}
    \label{fig:position_robustness}
    \vspace{-5mm}
\end{figure}

\begin{figure}[t]
    \centering
    \includegraphics[scale=0.32]{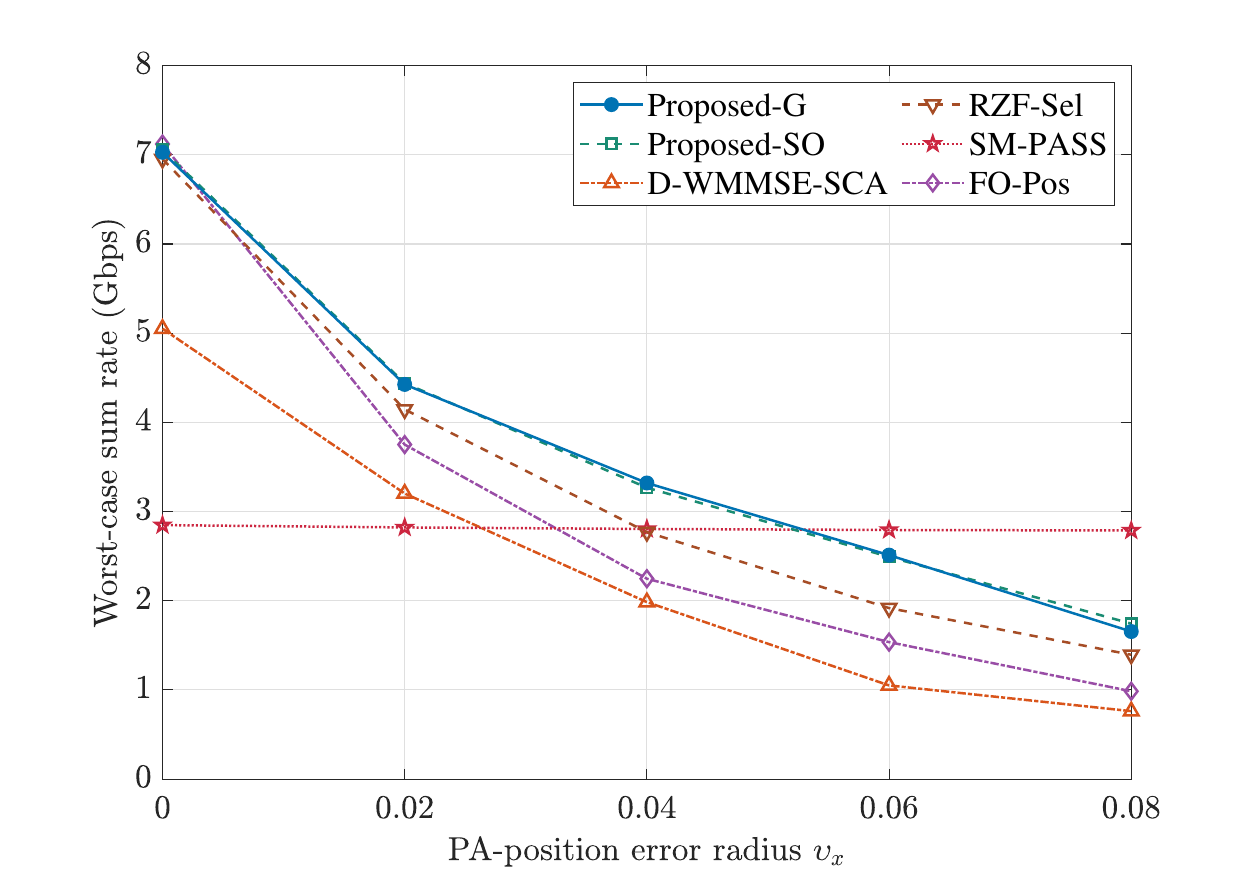}
    \caption{Worst-case sum rate versus the normalized
    PA positioning error radius for the proposed
    algorithms and benchmark schemes
    ($\epsilon_h=6\times10^{-6}$).}
    \label{fig:position_algorithms}
    \vspace{-5mm}
\end{figure}

\begin{figure}[t]
    \centering
    \includegraphics[scale=0.32]{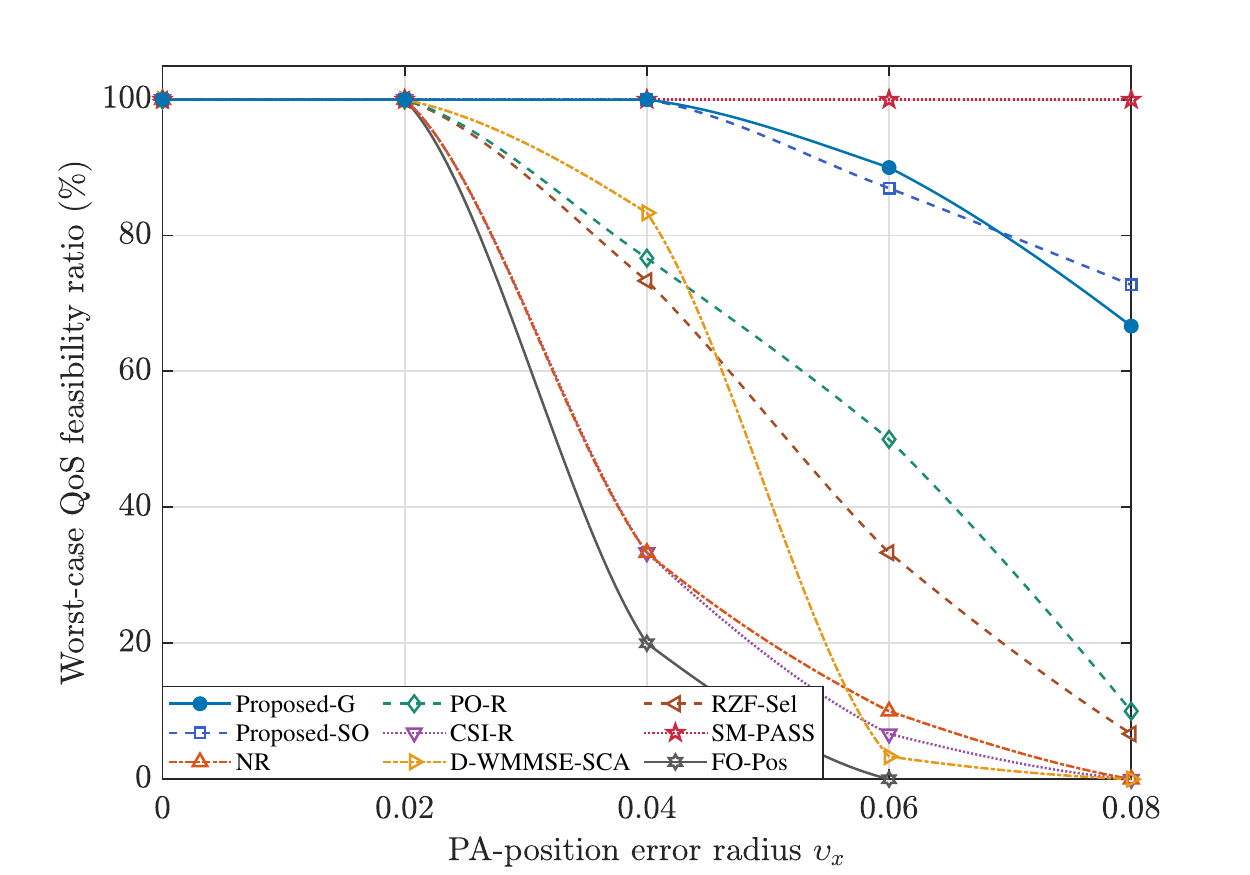}
    \caption{Worst-case QoS feasibility ratio versus
    the normalized PA positioning error radius for different
    schemes ($\epsilon_h=6\times10^{-6}$).}
    \label{fig:position_qos}
    \vspace{-5mm}
\end{figure}

Figure~\ref{fig:position_robustness} shows the
worst-case sum rate versus the PA positioning error radius for different error modeling schemes.
At $\upsilon_x=0.08$, Proposed-G achieves sum-rate
gains of approximately $55.5\%$ and $55.2\%$
over NR and CSI-R, respectively, demonstrating
the importance of accounting for PA positioning errors during optimization. Although PO-R achieves
a comparable sum rate and even slightly
outperforms Proposed-G at larger $\upsilon_x$,
neglecting CSI errors leads to more frequent
QoS violations, as will be shown in
Fig.~\ref{fig:position_qos}.

Figure~\ref{fig:position_algorithms} compares the worst-case sum rates of
the proposed algorithms and benchmark schemes versus the
PA positioning error radius. Proposed-G and Proposed-SO achieve comparable
performance and outperform the other multi-mode schemes for
$\upsilon_x>0$. Compared with SM-PASS, the proposed schemes achieve
significantly higher sum rates under small and moderate PA positioning errors, thanks to the spatial multiplexing gain of multi-mode
transmission. As $\upsilon_x$ increases, this gain gradually
diminishes, and SM-PASS achieves a higher sum rate when
$\upsilon_x\ge0.06$. This is because multi-mode transmission relies on
accurate channel phases to suppress inter-user interference, whose
residual increases with the PA positioning error, whereas SM-PASS is only mildly affected, as discussed for
Fig.~\ref{fig:csi_algorithms}. This reveals a tradeoff between
multiplexing gain and robustness, and indicates that sufficiently
accurate PA positioning is required to fully exploit multi-mode
transmission.

Figure~\ref{fig:position_qos} shows the worst-case
QoS feasibility ratio versus the PA positioning error radius. As $\upsilon_x$ increases, the feasibility ratios of most schemes decrease,
indicating that larger PA positioning errors make
it more difficult to satisfy the user QoS
constraints. The proposed methods outperform
all multi-mode benchmark schemes at large
$\upsilon_x$. In contrast, NR, CSI-R, and FO-Pos
experience a rapid degradation in QoS feasibility
as the PA positioning errors increase. Meanwhile, SM-PASS maintains full feasibility across all tested error levels, since its interference-free OFDMA transmission is only mildly
affected by the PA positioning errors, as discussed for Fig.~\ref{fig:position_algorithms}. However, this robustness comes at the cost of a
significantly lower sum rate under small and moderate PA positioning
errors, which reflects the tradeoff discussed above.

\begin{figure}[t]
    \centering
    \includegraphics[scale=0.32]{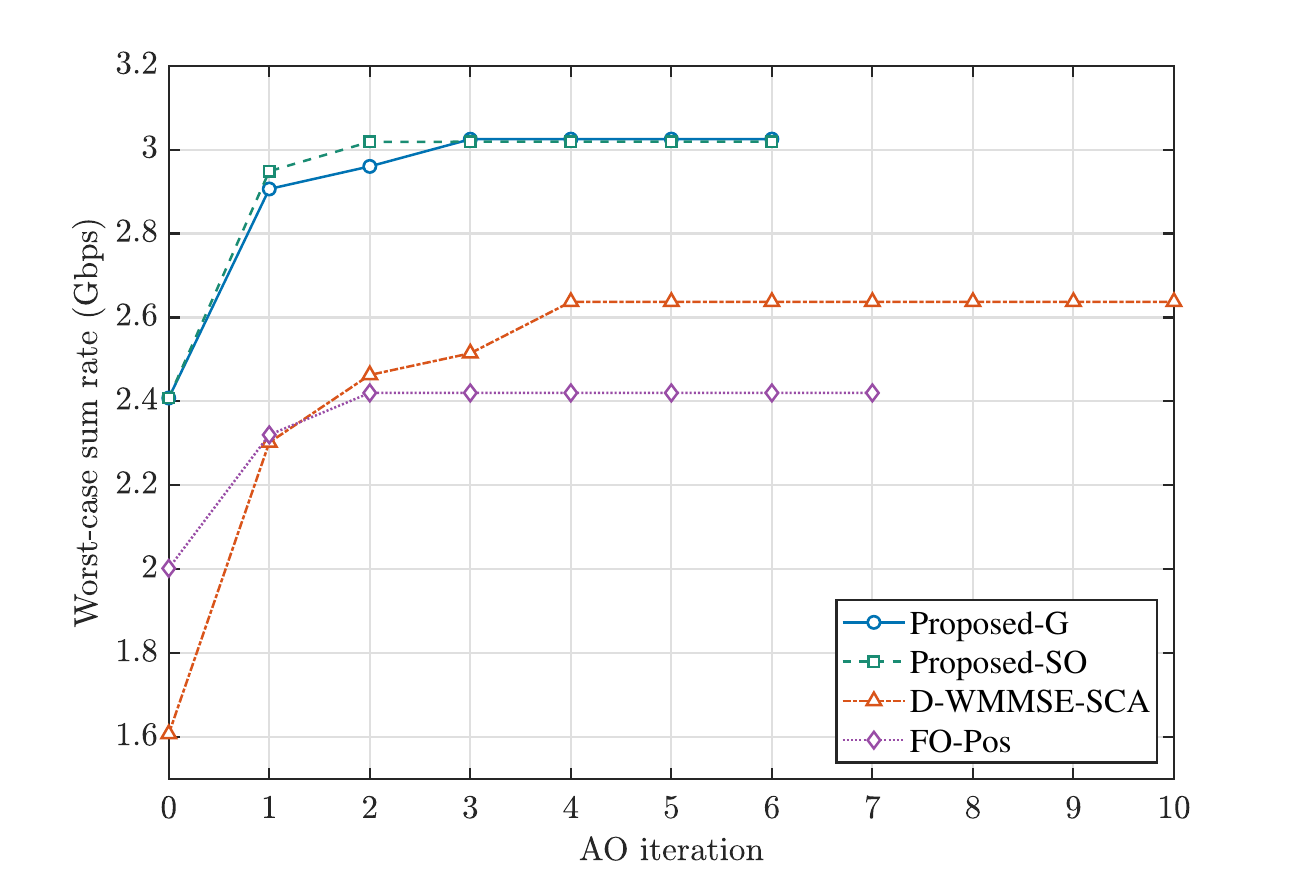}
    \caption{Worst-case sum rate versus the number of AO iterations
    for the proposed algorithms and benchmark schemes
    ($\epsilon_h=6\times10^{-6}$, $\upsilon_x=0.04$).}
    \label{fig:convergence}
    \vspace{-5mm}
\end{figure}

Figure~\ref{fig:convergence} illustrates the
convergence behavior of different schemes from a single realization. Both Proposed-G and Proposed-SO approach
stable worst-case sum rates within a few
AO iterations and achieve higher final
sum rates than D-WMMSE-SCA and FO-Pos.
These results demonstrate the effectiveness of the proposed algorithms.

\section{Conclusion}
\label{sec:conclusion}
This paper studied robust joint beamforming and PA-position optimization
for MPA-based OFDM systems under both CSI errors and PA positioning errors.
Based on a CME-based wideband channel model, it was shown that the two
errors, although arising from independent sources, are coupled through
the effective channel. To handle this coupling, the worst-case CSI error
was obtained exactly for each PA-position realization, while the
PA positioning errors were represented by grid-based and second-order
approximation-based scenarios, leading to a robust AO algorithm with
KKT-based updates. Numerical results showed that jointly handling both
errors is essential. Neglecting PA positioning errors led to a significant
loss in the worst-case sum rate, while neglecting CSI errors caused
frequent QoS violations despite a comparable sum rate. Compared with
conventional robust optimization schemes, the proposed designs achieved
higher worst-case sum rates and maintained high QoS feasibility under
various CSI and PA positioning error levels. 

\FloatBarrier
\appendices
\section{Proof of Proposition 1}
\label{app:proof_adversary}
\setcounter{equation}{0}
\numberwithin{equation}{section}
\renewcommand{\theequation}{\thesection.\arabic{equation}}
For fixed $\mathbf\Omega_p$, define
{\small
\begin{align}
    S_{k,p}(\mathbf q_{k,p};\mathbf\Omega_p)
    &\triangleq
    \left|
        \mu_{k,k,p}(\mathbf\Omega_p)
        +
        \mathbf q_{k,p}^{H}
        \boldsymbol{\nu}_{k,k,p}(\mathbf\Omega_p)
    \right|^2,
    \label{eq:app_N}
    \\
    D_{k,p}(\mathbf q_{k,p};\mathbf\Omega_p)
    &\triangleq
    1+
   \sum_{\substack{j\in\mathcal K\\ j\neq k}}
    \left|
        \mu_{k,j,p}(\mathbf\Omega_p)
        +
        \mathbf q_{k,p}^{H}
        \boldsymbol{\nu}_{k,j,p}(\mathbf\Omega_p)
    \right|^2 .
    \label{eq:app_D}
\end{align}}
\hspace*{-\parindent}~Since $D_{k,p}(\mathbf q_{k,p};\mathbf\Omega_p)\ge1$
for every feasible $\mathbf q_{k,p}$, by the definition of
$\Gamma_{k,p}^{\rm wc}(\mathbf\Omega_p)$,
\begin{equation}
    \Gamma_{k,p}^{\rm wc}(\mathbf\Omega_p)
    =
    \min_{\|\mathbf q_{k,p}\|_2\le1}
    \frac{
        S_{k,p}(\mathbf q_{k,p};\mathbf\Omega_p)
    }{
        D_{k,p}(\mathbf q_{k,p};\mathbf\Omega_p)
    }.
    \label{eq:app_gamma}
\end{equation}
Hence, for every feasible $\mathbf q_{k,p}$,
\begin{equation}
\begin{aligned}
    S_{k,p}(\mathbf q_{k,p};\mathbf\Omega_p)
    -
    \Gamma_{k,p}^{\rm wc}(\mathbf\Omega_p)
    D_{k,p}(\mathbf q_{k,p};\mathbf\Omega_p)
    \ge0,
\end{aligned}
\label{eq:app_nonnegative}
\end{equation}
where equality is attained at an optimizer of
\eqref{eq:app_gamma}. Therefore,
\begin{equation}
    \phi_{k,p}
    \left(
        \Gamma_{k,p}^{\rm wc}(\mathbf\Omega_p);
        \mathbf\Omega_p
    \right)
    =0.
    \label{eq:app_root}
\end{equation}
Moreover, for any $\tau_2>\tau_1\ge0$, let
$\mathbf q_{k,p}^{(1)}$ be an optimizer of
$\phi_{k,p}(\tau_1;\mathbf\Omega_p)$. Then,
\begin{equation}
\begin{aligned}
    \phi_{k,p}(\tau_2;\mathbf\Omega_p)
    &\le
    S_{k,p}(\mathbf q_{k,p}^{(1)};\mathbf\Omega_p)
    -
    \tau_2
    D_{k,p}(\mathbf q_{k,p}^{(1)};\mathbf\Omega_p)
    \\
    &=
    \phi_{k,p}(\tau_1;\mathbf\Omega_p)
    -
    (\tau_2-\tau_1)
    D_{k,p}(\mathbf q_{k,p}^{(1)};\mathbf\Omega_p)
    \\
    &\le
    \phi_{k,p}(\tau_1;\mathbf\Omega_p)
    -
    (\tau_2-\tau_1)<
    \phi_{k,p}(\tau_1;\mathbf\Omega_p),
\end{aligned}
\label{eq:app_strict_decrease}
\end{equation}
where the second inequality follows from
$D_{k,p}(\mathbf q_{k,p};\mathbf\Omega_p)\ge1$.
Hence, $\phi_{k,p}(\tau;\mathbf\Omega_p)$ is strictly
decreasing in $\tau$. Together with \eqref{eq:app_root},
this shows that
$\Gamma_{k,p}^{\rm wc}(\mathbf\Omega_p)$ is the unique
nonnegative root of
$\phi_{k,p}(\tau;\mathbf\Omega_p)$.

For any fixed $\tau\ge0$, expanding $ S_{k,p}(\mathbf q_{k,p};\mathbf\Omega_p)-\tau D_{k,p}(\mathbf q_{k,p};\mathbf\Omega_p)$
yields
\par{\small
\begin{equation}
\begin{aligned}
    &\mathbf q_{k,p}^{H}
    \mathbf L_{k,p}(\tau;\mathbf\Omega_p)
    \mathbf q_{k,p}
    +
    2{\rm Re}
    \left\{
        \mathbf q_{k,p}^{H}
        \mathbf b_{k,p}(\tau;\mathbf\Omega_p)
    \right\}
    +
    \vartheta_{k,p}(\tau;\mathbf\Omega_p),
\end{aligned}
\label{eq:app_quadratic}
\end{equation}}
\hspace*{-\parindent}~\,where
\begin{align}
    \mathbf L_{k,p}(\tau;\mathbf\Omega_p)
    &=
    \boldsymbol{\nu}_{k,k,p}(\mathbf\Omega_p)
    \boldsymbol{\nu}_{k,k,p}^{H}(\mathbf\Omega_p)
    \nonumber\\
    &\quad-
    \tau
    \!\!\!\!\sum_{\substack{j\in\mathcal K, j\neq k}}
    \boldsymbol{\nu}_{k,j,p}(\mathbf\Omega_p)
    \boldsymbol{\nu}_{k,j,p}^{H}(\mathbf\Omega_p),
    \label{eq:app_L}
    \\
    \mathbf b_{k,p}(\tau;\mathbf\Omega_p)
    &=
    \mu_{k,k,p}^{*}(\mathbf\Omega_p)
    \boldsymbol{\nu}_{k,k,p}(\mathbf\Omega_p)
    \nonumber\\
    &\quad-
    \tau
    \!\!\!\!\sum_{\substack{j\in\mathcal K, j\neq k}}
    \mu_{k,j,p}^{*}(\mathbf\Omega_p)
    \boldsymbol{\nu}_{k,j,p}(\mathbf\Omega_p),
    \label{eq:app_b}
    \\
    \vartheta_{k,p}(\tau;\mathbf\Omega_p)
    &=
    |\mu_{k,k,p}(\mathbf\Omega_p)|^2
    \nonumber\\
    &\quad-
    \tau
    \big(
        1+
        \!\!\!\!\sum_{\substack{j\in\mathcal K, j\neq k}}
        |\mu_{k,j,p}(\mathbf\Omega_p)|^2
    \big).
    \label{eq:app_vartheta}
\end{align}
Thus, for fixed $\tau$, the inner minimization is a quadratic
minimization over the unit Euclidean ball, i.e., a complex
trust-region subproblem. Its low-dimensional structure is exploited as follows.

Define $ \mathcal V_{k,p}(\mathbf\Omega_p)
    \triangleq
    {\rm span}
    \left\{
        \boldsymbol{\nu}_{k,1,p}(\mathbf\Omega_p),
        \ldots,
        \boldsymbol{\nu}_{k,K,p}(\mathbf\Omega_p)
    \right\}.$
Any feasible $\mathbf q_{k,p}$ can be decomposed as
$\mathbf q_{k,p}=\mathbf q_{k,p}^{\rm proj}+\mathbf q_{k,p}^{\rm res}$,
where $\mathbf q_{k,p}^{\rm proj}\in\mathcal V_{k,p}(\mathbf\Omega_p)$
and $\mathbf q_{k,p}^{\rm res}\perp\mathcal V_{k,p}(\mathbf\Omega_p)$.
It follows that $\mathbf q_{k,p}^{H}\boldsymbol{\nu}_{k,j,p}(\mathbf\Omega_p)
=(\mathbf q_{k,p}^{\rm proj})^H\boldsymbol{\nu}_{k,j,p}(\mathbf\Omega_p)$,
$\forall j\in\mathcal K$.
Therefore, both
$S_{k,p}(\mathbf q_{k,p};\mathbf\Omega_p)$ and
$D_{k,p}(\mathbf q_{k,p};\mathbf\Omega_p)$
depend only on $\mathbf q_{k,p}^{\rm proj}$.
Since $\|\mathbf q_{k,p}^{\rm proj}\|_2\le\|\mathbf q_{k,p}\|_2$, removing $\mathbf q_{k,p}^{\rm res}$ leaves the
objective unchanged while preserving feasibility. An optimal
CSI-error vector can therefore always be chosen from
$\mathcal V_{k,p}(\mathbf\Omega_p)$, whose dimension
$d_{k,p}$ satisfies $d_{k,p}\le\min\{N,M,K\}$.

Let
$\mathbf V_{k,p}(\mathbf\Omega_p)
\in\mathbb C^{N\times d_{k,p}}$
contain an orthonormal basis of
$\mathcal V_{k,p}(\mathbf\Omega_p)$.
Then, any vector in
$\mathcal V_{k,p}(\mathbf\Omega_p)$ can be represented as
$\mathbf q_{k,p}=\mathbf V_{k,p}(\mathbf\Omega_p)\tilde{\mathbf q}_{k,p}$.
Since
$\mathbf V_{k,p}^{H}(\mathbf\Omega_p)
\mathbf V_{k,p}(\mathbf\Omega_p)
=
\mathbf I_{d_{k,p}}$,
we have
$\|\mathbf q_{k,p}\|_2
=
\|\tilde{\mathbf q}_{k,p}\|_2$.
Therefore, restricting the search to
$\mathcal V_{k,p}(\mathbf\Omega_p)$ does not change the
optimal value, and the constraint
$\|\mathbf q_{k,p}\|_2\le1$
becomes
$\|\tilde{\mathbf q}_{k,p}\|_2\le1$.

Substituting $\mathbf q_{k,p}=\mathbf V_{k,p}(\mathbf\Omega_p)\tilde{\mathbf q}_{k,p}$ into
\eqref{eq:app_quadratic}, the inner minimization is
equivalently reduced to
\begin{equation}
\begin{aligned}
    \min_{\|\tilde{\mathbf q}_{k,p}\|_2\le1}
    \quad&
    \tilde{\mathbf q}_{k,p}^{H}
    \widetilde{\mathbf L}_{k,p}
    (\tau;\mathbf\Omega_p)
    \tilde{\mathbf q}_{k,p}
    \\
    &+
    2{\rm Re}
    \left\{
        \tilde{\mathbf q}_{k,p}^{H}
        \widetilde{\mathbf b}_{k,p}
        (\tau;\mathbf\Omega_p)
    \right\}
    +
    \vartheta_{k,p}(\tau;\mathbf\Omega_p),
\end{aligned}
\label{eq:app_reduced_trs}
\end{equation}
where $\widetilde{\mathbf L}_{k,p}(\tau;\mathbf\Omega_p)=
\mathbf V_{k,p}^{H}(\mathbf\Omega_p)\mathbf L_{k,p}(\tau;\mathbf\Omega_p)
\mathbf V_{k,p}(\mathbf\Omega_p)$ and
$\widetilde{\mathbf b}_{k,p}(\tau;\mathbf\Omega_p)=
\mathbf V_{k,p}^{H}(\mathbf\Omega_p)\mathbf b_{k,p}(\tau;\mathbf\Omega_p)$.
Problem~\eqref{eq:app_reduced_trs} is a standard complex
trust-region subproblem whose dimension is no larger than
$\min\{N,M,K\}$. Its globally optimal solution can be obtained
from the standard trust-region optimality conditions
\cite{more1983trust}.

Finally, the original fractional minimization is solved by the
Dinkelbach procedure \cite{dinkelbach1967fractional}. At iteration
$\ell$, for a given $\tau_{k,p}^{(\ell)}$, let
$\tilde{\mathbf q}_{k,p}^{(\ell)}$
denote a globally optimal solution of
\eqref{eq:app_reduced_trs}. The corresponding CSI-error vector
is recovered as $   \mathbf q_{k,p}^{(\ell)}=
    \mathbf V_{k,p}(\mathbf\Omega_p)
    \tilde{\mathbf q}_{k,p}^{(\ell)}.$
The parameter is then updated according to
\begin{equation}
    \tau_{k,p}^{(\ell+1)}
    =
    \frac{
        S_{k,p}
        \left(
            \mathbf q_{k,p}^{(\ell)};
            \mathbf\Omega_p
        \right)
    }{
        D_{k,p}
        \left(
            \mathbf q_{k,p}^{(\ell)};
            \mathbf\Omega_p
        \right)
    }.
    \label{eq:app_dinkelbach_update}
\end{equation}
The iterations terminate when $ \left|
        \phi_{k,p}
        \left(
            \tau_{k,p}^{(\ell)};
            \mathbf\Omega_p
        \right)
    \right|
    \le
    \varepsilon_{\rm D},$
where $\varepsilon_{\rm D}>0$ is a prescribed tolerance.
Since $\phi_{k,p}(\tau;\mathbf\Omega_p)$ is strictly decreasing and has
the unique nonnegative root $\Gamma_{k,p}^{\rm wc}(\mathbf\Omega_p)$, the
sequence $\{\tau_{k,p}^{(\ell)}\}$ converges to
$\Gamma_{k,p}^{\rm wc}(\mathbf\Omega_p)$~\cite{dinkelbach1967fractional}.
Upon convergence, the associated $\mathbf q_{k,p}$ satisfies
$S_{k,p}(\mathbf q_{k,p};\mathbf\Omega_p)
=\Gamma_{k,p}^{\rm wc}(\mathbf\Omega_p)
D_{k,p}(\mathbf q_{k,p};\mathbf\Omega_p)$, and is thus a globally optimal
CSI-error vector of \eqref{eq:wc_sinr}, which completes the proof.

\section{Proof of Proposition 2}
\label{app:proof_w_update}
The Lagrangian function of Problem~\eqref{prob:w_update} can be written as
{
\begin{align}
    \mathcal L
    =&\;
    t_{\rm w}
    +
    \sum_{s\in\mathcal S}
    \theta_s
    \big(
    \sum_{k\in\mathcal K}
    \sum_{p\in\mathcal P}
    \xi_{k,p}^{(s, r)}
    -
    t_{\rm w}
    \big)
    \notag\\
    &+
    \sum_{s\in\mathcal S}
    \sum_{k\in\mathcal K}
    \lambda_{k,s}
    \big(
    \sum_{p\in\mathcal P}
    \xi_{k,p}^{(s, r)}
    -
    c_k
    \big)
    \notag\\
    &+
    \sum_{p\in\mathcal P}
    \rho_p
    \left(
    \|\mathbf W_p\|_F^2
    -
    P_{p,\max}
    \right).
    \label{eq:app_w_lagrangian}
\end{align}}

Collecting the terms associated with the epigraph variable
$t_{\rm w}$, the Lagrangian contains
$\left(1-\sum_{s\in\mathcal S}\theta_s\right)t_{\rm w}$.
Since $t_{\rm w}$ is unconstrained, stationarity with respect to $t_{\rm w}$
requires $\sum_{s\in\mathcal S}\theta_s=1$. Hence, the dual variables
associated with the epigraph constraints lie on the unit
simplex.

For fixed $u_{k,p}^{(s,r)}$ and $v_{k,p}^{(s,r)}$, each
$\xi_{k,p}^{(s,r)}$ in \eqref{eq:robust_mse}--\eqref{eq:robust_wmse} is a
convex quadratic function of $\mathbf W_p$.
Collecting all quadratic and linear terms in
\eqref{eq:app_w_lagrangian} and imposing stationarity with
respect to $\mathbf W_p^*$ yields
\begin{equation}
   (
    \mathbf Q_p^{\rm W}
    +
    \rho_p\mathbf I_M
    )
    \mathbf W_p
    =
    \mathbf Z_p^{\rm W},
    \label{eq:app_w_stationarity}
\end{equation}
where $\mathbf Q_p^{\rm W}$ and $\mathbf Z_p^{\rm W}$ are
defined in \eqref{eq:Qw_def}--\eqref{eq:zw_def}.

To solve the KKT system of Problem~\eqref{prob:w_update}, define
$\boldsymbol{\theta}\triangleq[\theta_1,\ldots,\theta_{|\mathcal S|}]^T$
and $\boldsymbol{\lambda}\triangleq
\{\lambda_{k,s}\}_{k\in\mathcal K,s\in\mathcal S}$. For fixed
$\boldsymbol{\theta}$ and $\boldsymbol{\lambda}$, $\mathbf Q_p^{\rm W}$
and $\mathbf Z_p^{\rm W}$ are constructed via
\eqref{eq:Qw_def}--\eqref{eq:zw_def}, and minimizing the Lagrangian
over $\mathbf W_p$ subject to the per-subcarrier power constraint
yields
\begin{equation}
    \mathbf W_p(\boldsymbol\theta,\boldsymbol\lambda)
    =
    \left(\mathbf Q_p^{\rm W}+\rho_p\mathbf I_M\right)^{-1}
    \mathbf Z_p^{\rm W}.
    \label{eq:app_W_rho}
\end{equation}
If $\mathbf Q_p^{\rm W}$ is singular and $\rho_p=0$, the minimum-norm
solution $\mathbf W_p(\boldsymbol\theta,\boldsymbol\lambda)
=(\mathbf Q_p^{\rm W})^{\dagger}\mathbf Z_p^{\rm W}$ is adopted, which
is well defined since the columns of $\mathbf Z_p^{\rm W}$ lie in the
column space of $\mathbf Q_p^{\rm W}$.
If the solution obtained with $\rho_p=0$ satisfies
$\|\mathbf W_p(\boldsymbol\theta,\boldsymbol\lambda)\|_F^2\le P_{p,\max}$,
the power constraint is inactive and complementary slackness gives
$\rho_p=0$. Otherwise, the power constraint is active and $\rho_p>0$ is
determined from
\begin{equation}
    \left\|
        \left(
            \mathbf Q_p^{\rm W}
            +
            \rho_p\mathbf I_M
        \right)^{-1}
        \mathbf Z_p^{\rm W}
    \right\|_F^2
    =
    P_{p,\max}.
    \label{eq:app_rho_bisection}
\end{equation}
Since the left-hand side of \eqref{eq:app_rho_bisection} is
non-increasing in $\rho_p$, $\rho_p$ can be efficiently obtained by
scalar bisection.

Since $\sum_{s\in\mathcal S}\theta_s=1$, the terms associated with $t_{\rm w}$
in \eqref{eq:app_w_lagrangian} vanish. Substituting
\eqref{eq:app_W_rho} into \eqref{eq:app_w_lagrangian} then yields the
dual function $\Psi(\boldsymbol\theta,\boldsymbol\lambda)$, which is
concave~\cite{vandenberghe2004convex}. Since the minimizer in
\eqref{eq:app_W_rho} is unique when
$\mathbf Q_p^{\rm W}+\rho_p\mathbf I_M\succ\mathbf 0$,
$\Psi(\boldsymbol\theta,\boldsymbol\lambda)$ is differentiable, and its
gradient is given by the constraint functions evaluated at
$\mathbf W(\boldsymbol\theta,\boldsymbol\lambda)$, which is given by
\begin{equation}
\begin{aligned}
    \frac{\partial \Psi}{\partial\theta_s}
    &=
    \sum_{k\in\mathcal K}\sum_{p\in\mathcal P}
    \xi_{k,p}^{(s,r)}
    \big(\mathbf W_p(\boldsymbol\theta,\boldsymbol\lambda)\big),
    \\
    \frac{\partial \Psi}{\partial\lambda_{k,s}}
    &=
    \sum_{p\in\mathcal P}
    \xi_{k,p}^{(s,r)}
    \big(\mathbf W_p(\boldsymbol\theta,\boldsymbol\lambda)\big)
    -c_k .
\end{aligned}
\label{eq:app_dual_gradient}
\end{equation}
If $\mathbf Q_p^{\rm W}+\rho_p\mathbf I_M$ is singular,
\eqref{eq:app_dual_gradient} provides a supergradient of
$\Psi(\boldsymbol\theta,\boldsymbol\lambda)$.
Accordingly, the dual variables are obtained by solving
\begin{equation}
\begin{aligned}
&\max_{\boldsymbol\theta,\boldsymbol\lambda}    \Psi(\boldsymbol\theta,\boldsymbol\lambda) \\
    &~~\mathrm{s.t.}\;
    \boldsymbol\theta\succeq\mathbf 0,\;
    \sum_{s\in\mathcal S}\theta_s=1,\;
    \lambda_{k,s}\ge0,\;
    \forall k\in\mathcal K,\, s\in\mathcal S,
    \label{eq:app_dual_problem}
    \end{aligned}
\end{equation}
which can be solved by sequential quadratic programming (SQP)
method~\cite{nocedal2006numerical}. The procedure
terminates when the relative duality gap and the primal constraint
violation fall below prescribed tolerances. Denote the beamformers
recovered at the resulting dual variables by $\mathbf W^{(r,\star)}$.
Since Problem~\eqref{prob:w_update} is convex and satisfies Slater's
condition, strong duality holds. Hence, $\mathbf W^{(r,\star)}$
satisfies \eqref{eq:app_w_stationarity} together with primal
feasibility, dual feasibility, and complementary slackness, and is therefore an optimal solution of Problem~\eqref{prob:w_update}, which completes the proof.

\section{Proof of Proposition 3}
\label{app:proof_x_update}

The Lagrangian of Problem~\eqref{prob:x_update} is
\par{\small
\begin{equation}
\begin{aligned}
    \mathcal L_{\rm x}
    =&\;
    t_{\rm x}
    +
    \sum_{s\in\mathcal S}
    \chi_s
    \left(
        \sum_{k\in\mathcal K}
        \sum_{p\in\mathcal P}
        \widetilde{\xi}_{k,p}^{\mathrm x,(s,r)}
        -
        t_{\rm x}
    \right)
    \\
    &+
    \sum_{k\in\mathcal K}
    \sum_{s\in\mathcal S}
    \eta_{k,s}
    \left(
        \sum_{p\in\mathcal P}
        \widetilde{\xi}_{k,p}^{\mathrm x,(s,r)}
        -
        c_k
    \right)
    \\
    &+
    \sum_{n=1}^{N}
   \varrho_n^{\mathrm{L}}
    \left(
        x_{\min}+\delta_x-\widehat x_n
    \right)+
    \sum_{n=1}^{N}
    \varrho_n^{\mathrm{U}}
    \left(
        \widehat x_n-x_{\max}+\delta_x
    \right)
    \\
    &+
    \sum_{n=1}^{N-1}
    \varphi_n
    \left(
        \Delta_{\min}
        +2\delta_x
        +\widehat x_n-\widehat x_{n+1}
    \right).
\end{aligned}
\label{eq:app_x_lagrangian}
\end{equation}}

Setting the derivative of $\mathcal L_{\rm x}$ with respect to
$t_{\rm x}$ to zero yields $\sum_{s\in\mathcal S}\chi_s=1$.

From \eqref{eq:x_wmse_surrogate}, the gradient of each
surrogate WMSE with respect to $\widehat{\mathbf x}$ is
\begin{equation}
\begin{aligned}
    \nabla_{\widehat{\mathbf x}}
    \widetilde{\xi}_{k,p}^{\mathrm x,(s,r)}
    =
    \nabla_{\widehat{\mathbf x}}
    \xi_{k,p}^{\mathrm x,(s,r)}
    +
    \left(
        \mathbf T_{k,p}^{(s,r)}
        +
        \iota^{(r)}\mathbf I_N
    \right)
    \left(
        \widehat{\mathbf x}
        -
        \widehat{\mathbf x}^{(r)}
    \right).
\end{aligned}
\label{eq:app_x_surrogate_gradient}
\end{equation}

Using \eqref{eq:app_x_surrogate_gradient}, the stationarity condition
with respect to $\widehat{\mathbf x}$ is
\par{\small
\begin{equation}
\begin{aligned}
    &\sum_{s\in\mathcal S}
    \sum_{k\in\mathcal K}
    (\chi_s+\eta_{k,s})
    \sum_{p\in\mathcal P}
    \Big[
        \nabla_{\widehat{\mathbf x}}
        \xi_{k,p}^{\mathrm x,(s,r)}
    \\
    &+
        \big(
            \mathbf T_{k,p}^{(s,r)}
            +
            \iota^{(r)}\mathbf I_N
        \big)
        \big(
            \widehat{\mathbf x}
            -
            \widehat{\mathbf x}^{(r)}
        \big)
    \Big]
    -
    \sum_{n=1}^{N}
   \varrho_n^{\mathrm{L}}\mathbf e_n
    \\
    &+
    \sum_{n=1}^{N}
    \varrho_n^{\mathrm{U}}\mathbf e_n
    +
    \sum_{n=1}^{N-1}
    \varphi_n
    \left(
        \mathbf e_n-\mathbf e_{n+1}
    \right)
    =
    \mathbf 0.
\end{aligned}
\label{eq:app_x_stationarity}
\end{equation}}

Evaluating \eqref{eq:app_x_stationarity} at
$\widehat{\mathbf x}
=
\widehat{\mathbf x}_{\rm cand}^{(r+1)}$
and collecting terms according to \eqref{eq:x_Xi_def} and \eqref{eq:x_zeta_def} yields \eqref{eq:x_kkt_stationarity}.

Since $\sum_{s\in\mathcal S}\chi_s=1$, $\chi_s\ge0$, $\eta_{k,s}\ge0$,
and $\mathbf T_{k,p}^{(s,r)}+\iota^{(r)}\mathbf I_N\succ\mathbf 0$, it
follows from \eqref{eq:x_Xi_def} that
$\mathbf\Xi^{\mathrm x}\succeq KP\iota^{(r)}\mathbf I_N\succ\mathbf 0$.
Hence, $\mathbf\Xi^{\mathrm x}$ is invertible, and
\eqref{eq:x_closed_form} follows from
\eqref{eq:x_kkt_stationarity}.

To solve the KKT conditions, an active-set Newton method is
adopted~\cite{nocedal2006numerical}. Since the complementary slackness
conditions require identifying the constraints that are active at the
solution, the multipliers of the inactive constraints are set to zero,
and the stationarity conditions, together with the active constraints,
are solved for $\widehat{\mathbf x}$, $t_{\rm x}$, and the active
multipliers by Newton's method. The active set is then updated by
removing the constraints with negative multipliers and adding the
violated constraints. Upon convergence, the obtained solution satisfies
\eqref{eq:x_kkt_stationarity} together with primal feasibility,
dual feasibility, and complementary slackness. Since
Problem~\eqref{prob:x_update} is a convex QCQP and satisfies Slater's
condition, these KKT conditions are necessary and sufficient for the optimality of Problem~\eqref{prob:x_update}, which completes the proof.

\bibliographystyle{IEEEtran}
\bibliography{ref}

\end{document}